\RequirePackage{fix-cm} 
\documentclass[a4paper, twoside, reqno, 12pt]{amsart}
\usepackage{fixltx2e}   

\usepackage{etex}

\usepackage[latin1]{inputenc}
\usepackage[T1]{fontenc}

\usepackage{esint}
\usepackage{dsfont}
\usepackage{xspace}
\usepackage{amsgen}
\usepackage{amsthm}
\usepackage{amssymb}
\usepackage{amsmath}
\usepackage{wasysym}
\usepackage{upgreek}
\usepackage{amsfonts}
\usepackage{stmaryrd}
\usepackage{mathtools}

\usepackage[mathcal, mathscr]{euscript}

\usepackage{mathrsfs}
\DeclareMathAlphabet{\mathscrbf}{OMS}{mdugm}{b}{n}

\usepackage{algpseudocode, algorithm, algorithmicx}

\usepackage{multicol}
\usepackage{multirow}

\usepackage{a4wide}

\usepackage[dvipsnames, table]{xcolor}
\definecolor{bckg}{RGB}{20.8, 20.8, 20.8}
\definecolor{oneblue}{rgb}{0.0, 0.0, 0.85}
\definecolor{Lightblue}{RGB}{214, 214, 214}
\definecolor{bluepigment}{rgb}{0.2, 0.2, 0.6}
\definecolor{charcoal}{rgb}{0.21, 0.27, 0.31}
\definecolor{denimblue}{rgb}{0.08, 0.38, 0.74}
\definecolor{Lightgray}{rgb}{0.89, 0.89, 0.89}
\definecolor{darkgrey}{rgb}{0.273, 0.281, 0.30}
\definecolor{darkelectricblue}{rgb}{0.33, 0.41, 0.47}

\usepackage[sort&compress, comma, square, numbers]{natbib}

\usepackage{psfrag}
\usepackage{graphicx}
\usepackage{subfigure}
\usepackage{morefloats}
\usepackage{indentfirst}

\usepackage{acronym}
\usepackage{microtype}
\usepackage[labelsep=period,%
            labelfont={bf,sf,color=bluepigment},%
            justification=raggedright]{caption}

\usepackage[perpage, symbol]{footmisc}

\usepackage[usenames, pdf]{pstricks}
\usepackage{epsfig}
\usepackage{pst-grad} 
\usepackage{pst-plot} 

\usepackage{url}
\usepackage[colorlinks,
           urlcolor=oneblue,
           linkcolor=denimblue,
           citecolor=NavyBlue,
           bookmarksopen=false,
           pdfpagemode=UseNone,
           pagebackref]{hyperref}

\usepackage[explicit]{titlesec}

\titleformat{\section}[block]
  {\color{NavyBlue}\Large\sffamily\bfseries}
  {}
  {0.0em}
  {\colorbox{bckg!5}{\strut\parbox{\dimexpr\linewidth-2\fboxsep\relax}{\thesection. #1}}}
  [\vspace*{0.33em}]

\titleformat{name=\section,numberless}[block]
  {\color{NavyBlue}\Large\sffamily\bfseries}
  {}
  {0.0em}
  {\colorbox{bckg!5}{\strut\parbox{\dimexpr\linewidth-2\fboxsep\relax}{#1}}}
  [\vspace*{0.33em}]

\titleformat{\subsection}
  {\color{NavyBlue}\large\sffamily\bfseries}
  {}
  {0.0em}
  {\colorbox{bckg!5}{\parbox{\dimexpr\linewidth-2\fboxsep\relax}{\thesubsection. #1}}}
  [\vspace*{0.33em}]

\titleformat{name=\subsection,numberless}
  {\color{NavyBlue}\Large\sffamily\bfseries}
  {}
  {0em}
  {\colorbox{bckg!5}{\parbox{\dimexpr\linewidth-2\fboxsep\relax}{#1}}}
  [\vspace*{0.33em}]

\titleformat{\subsubsection}
  {\color{bluepigment}\sffamily\normalsize\bfseries}
  {\thesubsubsection}
  {0.5em}
  {#1}
  [\vspace*{0.33em}]

\titleformat{\paragraph}[runin]
  {\color{bluepigment}\sffamily\small\bfseries}
  {}
  {0em}
  {#1}

\titlespacing{\section}{0.0em}{1.5em plus 2pt minus 2pt}%
{1.0em plus 2pt minus 2pt}[0em]
\titlespacing{\subsection}{0.5em}{1.5em plus 2pt minus 2pt}%
{1.0em}[0em]
\titlespacing{\subsubsection}{0.5em}{1.5em plus 2pt minus 2pt}%
{1.0em plus 2pt minus 2pt}[0em]

\usepackage{titletoc}

\contentsmargin{0.5em}
\newlength{\tocsep} 
\titlecontents{section}[\tocsep]
  {\addvspace{10pt}\bfseries\sffamily}
  {\contentslabel[\thecontentslabel]{\tocsep}}
  {}
  {\ \titlerule*[0.75pc]{.}\ \thecontentspage}
  []
\titlecontents{subsection}[\tocsep]
  {\addvspace{8pt}\sffamily}
  {\contentslabel[\thecontentslabel]{\tocsep}}
  {}
  {\ \titlerule*[0.5pc]{.}\ \thecontentspage}
  []
\titlecontents*{subsubsection}[\tocsep]
  {\addvspace{2pt}\footnotesize\sffamily}
  {}
  {}
  {\ \titlerule*[0.35pc]{.}\ \thecontentspage}
  [\\*]

\makeatletter
\def\@setauthors{%
  \begingroup
  \def\thanks{\protect\thanks@warning}%
  \trivlist
  \centering\footnotesize \@topsep30\p@\relax
  \advance\@topsep by -\baselineskip
  \item\relax
  \author@andify\authors
  \def\\{\protect\linebreak}%
  \textsc{\normalsize\textcolor{darkelectricblue}{\authors}}%
  \ifx\@empty\contribs
  \else
    ,\penalty-3 \space \@setcontribs
    \@closetoccontribs
  \fi
  \endtrivlist
  \endgroup
}
\def\@settitle{\begin{center}%
  \baselineskip14\p@\relax
    \bfseries
    \textsc{\Large\textcolor{charcoal}{\@title}}
  \end{center}%
}
\makeatother

\usepackage{enumitem}
\setlist[description]{%
  topsep=30pt,               
  itemsep=5pt,               
  font={\bfseries\sffamily\color{NavyBlue}}, 
}

\usepackage{fancyhdr}
\usepackage{lastpage}

\newcommand*\Title{\textcolor{bluepigment}{Wave dynamics on networks}}
\newcommand*\Authors{\textcolor{bluepigment}{D.~Dutykh \& J.-G.~Caputo}}
\newcommand*{\plogo}{\textcolor{gray}{{\texttt{arXiv.org} / \textsc{hal}}}} 

\numberwithin{equation}{section}

\newtheorem{remark}{Remark}

\newcommand{\C}{\mathds{C}}
\newcommand{\I}{\mathds{I}}
\newcommand{\M}{\mathds{M}}
\newcommand{\N}{\mathds{N}}
\newcommand{\R}{\mathds{R}}
\newcommand{\Z}{\mathds{Z}}
\newcommand{\Ee}{\mathds{E}}
\newcommand{\D}{\mathcal{D}}
\newcommand{\E}{\mathcal{E}}

\newcommand{\ud}{\mathrm{d}}

\newcommand{\ue}{\mathrm{e}}
\newcommand{\Tt}{\mathbb{T}}
\newcommand{\T}{\mathcal{T}}
\newcommand{\Y}{\mathcal{Y}}

\newcommand{\In}{\mathcal{A}}

\renewcommand{\u}{\mathbf{u}}
\newcommand{\Mo}{\mathcal{M}}
\renewcommand{\P}{\mathcal{P}}

\renewcommand{\O}{\mathcal{O}}
\renewcommand{\H}{\mathcal{H}}
\renewcommand{\L}{\mathcal{L}}
\renewcommand{\S}{\mathcal{S}}
\newcommand{\n}{\boldsymbol{n}}
\newcommand{\x}{\boldsymbol{x}}
\renewcommand{\alpha}{\upalpha}
\newcommand{\z}{{\boldsymbol{z}}}
\newcommand{\J}{{\boldsymbol{J}}}
\renewcommand{\a}{{\boldsymbol{a}}}
\renewcommand{\b}{{\boldsymbol{b}}}
\renewcommand{\v}{{\boldsymbol{v}}}

\newcommand{\vd}{{\boldsymbol{\delta}}}

\renewcommand{\geq}{\geqslant}

\newcommand{\ie}{\emph{i.e.}\xspace}
\newcommand{\eg}{\emph{e.g.}\xspace}

\newcommand{\Mat}{\mathrm{Mat}\,}

\newcommand{\scal}{\boldsymbol{\cdot}}
\newcommand{\grad}{\boldsymbol{\nabla}}

\newcommand{\abs}[1]{\lvert\, #1\, \rvert}
\newcommand{\diag}{\mathop{\mathrm{diag}}}

\newcommand{\od}[2]{\frac{\mathrm{d}\/ #1}{\mathrm{d}\/#2}}

\newcommand{\eqdef}{\mathop{\stackrel{\,\mathrm{def}}{:=}\,}}

\newcommand{\half}{{\textstyle{1\over2}}}
\newcommand{\third}{{\textstyle{1\over3}}}

\usepackage{acronym}
\acrodef{sg}[sG]{sine--\textsc{Gordon}}
\acrodef{nswe}[NSWE]{Nonlinear Shallow Water Equations}

\begin{document}

\title[\Title]{Wave dynamics on networks: method and application to the sine--Gordon equation}

\author[D.~Dutykh]{Denys Dutykh$^*$}
\address{\textbf{D.~Dutykh:} LAMA, UMR 5127 CNRS, Universit\'e Savoie Mont Blanc, Campus Scientifique, F-73376 Le Bourget-du-Lac Cedex, France and Univ. Grenoble Alpes, Univ. Savoie Mont Blanc, CNRS, LAMA, 73000 Chamb\'ery, France}
\email{Denys.Dutykh@univ-savoie.fr}
\urladdr{http://www.denys-dutykh.com/}
\thanks{$^*$ Corresponding author}

\author[J.-G.~Caputo]{Jean-Guy Caputo}
\address{\textbf{J.-G.~Caputo:} Laboratoire de Math\'ematiques, INSA de Rouen, BP 8, Avenue de l'Universit\'e, Saint-Etienne du Rouvray, F-76801 France}
\email{caputo@insa-rouen.fr}
\urladdr{https://sites.google.com/site/jeanguycaputo/}



\begin{titlepage}
\setcounter{page}{1}
\thispagestyle{empty} 
\noindent
{\Large Denys \textsc{Dutykh}}\\
{\it\textcolor{gray}{CNRS, Universit\'e Savoie Mont Blanc, France}}\\[0.02\textheight]
{\Large Jean-Guy \textsc{Caputo}}\\
{\it\textcolor{gray}{INSA de Rouen, France}}\\[0.16\textheight]

\vspace*{0.75em}

\colorbox{Lightblue}{
  \parbox[t]{1.0\textwidth}{
    \centering\huge\sc
    \vspace*{0.7cm}
    
    \textcolor{bluepigment}{Wave dynamics on networks: method and application to the sine--Gordon equation}

    \vspace*{0.7cm}
  }
}

\vfill 

\raggedleft     
{\large \plogo} 
\end{titlepage}


\newpage
\thispagestyle{empty} 
\par\vspace*{\fill}   
\begin{flushright} 
{\textcolor{denimblue}{\textsc{Last modified:}} \today}
\end{flushright}


\newpage
\maketitle
\thispagestyle{empty}


\begin{abstract}

We consider a scalar \textsc{Hamiltonian} nonlinear wave equation formulated on networks; this is a non standard problem because these domains are not locally homeomorphic to any subset of the \textsc{Euclidean} space. More precisely, we assume each edge to be a 1D uniform line with end points identified with graph vertices. The interface conditions at these vertices are introduced and justified using conservation laws and an homothetic argument. We present a detailed methodology based on a symplectic finite difference scheme together with a special treatment at the junctions to solve the problem and apply it to the \acl{sg} equation. Numerical results on a simple graph containing four loops show the performance of the scheme for kinks and breathers initial conditions.


\bigskip
\noindent \textbf{\keywordsname:} Partial differential equations on networks; \textsc{Hamiltonian} partial differential equations; graph theory; sine--\textsc{Gordon} equation \\


\end{abstract}


\newpage
\tableofcontents
\thispagestyle{empty}


\newpage
\section{Introduction}

Currently there is a growing demand for modelling and understanding various flow problems on networks. A generic network is a (usually finite) set of points or simply \emph{vertices} immersed in an \textsc{Euclidean} space $\Ee^{\,2}$ or $\Ee^{\,3}$ (depending on the application in hands). Some of the points are connected by 1D segments (or more generally curves, which are homeomorphic to segments), called the \emph{edges}. Mathematically networks are formalized using graph theory \cite{Gould2012}. However, there is an important subtle difference with graph theory. Namely, in some applications the geometry of edges (\eg their length, shape, thickness) may matter, while in graph theory the only relevant information is the fact that two points are connected by an edge. Such sensitive applications include, for example blood flow modelling \cite{Vassilevskii2011}. Thus, a network combines in a single data structure the corresponding geometrical and topological information on vertices and edges. The flow is modeled with Partial Differential Equations (PDEs) because of the spatial dimension of the edges, as opposed to Ordinary Differential Equations (ODEs) in the standard case. For a general recent review of this topic, see \cite{Bressan2014}. This field continues to attract researchers from modelling, analysis, numerics, optimization and control theory \cite{Zuazua2013}.

One of the main difficulties of formulating evolution problems described by PDEs on networks (\ie graphs) consists in the fact that these objects are not manifolds. Recall that an $n-$dimensional manifold is such that each point has a neighborhood that is homeomorphic to $\R^{\,n}\,$ \cite{Abraham1988}. Nowadays, the formulation of \textsc{Hamiltonian} mechanics on manifolds does not pose any serious technical difficulties \cite{Arnold1997}. However, the modeling of various processes on networks, such as electric circuits, blood arteries, water-pipe supply needs the generalization of classical mechanics to non-manifolds like graphs and trees. Consider, for example, a $Y-$ or $T-$junction. This domain is a \emph{semi-algebraic set}, but not a \emph{manifold}. The difficulty comes from the branching point whose neighbourhood is not \emph{homeomorphic} to any \textsc{Euclidean} space $\Ee^{\,k}\,$. Therefore any composition of $Y-$junctions into, for example, a complex tree will not be a manifold either.

To our knowledge, \textsc{Hamiltonian} problems on non-manifolds have not been systematically studied. A notorious exception is the work \cite{Bibikov2009}, where wave scattering in the \textsc{Klein}--\textsc{Gordon}(--\textsc{Fok}) equation was investigated on a domain consisting of three semi-infinite straight lines having one common point. We can also mention the publication \cite{Bona2008} where the \textsc{Benjamin}--\textsc{Bona}--\textsc{Mahony} (BBM) equation was considered on a tree. The \acs{sg} equation on $Y-$shaped \textsc{Josephson} junctions was first considered in \cite{Nakajima1978, Nakajima1976}. The dynamics of kinks in $Y-$junctions was studied in \cite{Grunnet-Jepsen1993, Hattel1996}. However, these studies do not rely on any particular variational structure of the governing equation; the boundary conditions come from a particular tri-layer of superconducting films. The existence and stability of solitary waves `sitting' near the junction point was studied in \cite{Susanto2005}.

In the present study we consider the celebrated \acf{sg} equation which is a \textsc{Hamiltonian} and integrable PDE \cite{Takhtadzhyan1974, Dashen1974}. However, the integrability of the \acs{sg} equation is not compulsory for our purposes. In the developments presented below we will use the \textsc{Hamiltonian} and \textsc{Lagrangian} structures to determine the relevant conserved quantities and correct interface conditions at the vertices in order to construct an appropriate symplectic discretization. In the present study, we consider the discrete dynamics of \acs{sg} on $1-$D lattices assembled into a graph. The transition rules between the adjacent lattices at junction points follow from the discretization of the local conservation laws. This approach was already used in \cite{Bibikov2009, Caputo2014}. We expect that the limit of the lattice parameter $\Delta\ \to\ 0$ will provide us with the continuous version of the \textsc{Hamiltonian} mechanics on non-manifolds.

Our main result is a detailed methodology to solve \textsc{Hamiltonian} evolution equations on networks. We give this in full detail and explain which sections can be parallelized. For the case of the \acs{sg} or another nonlinear \textsc{Hamiltonian} equation, we justify the coupling conditions at the vertices of the network using a homothetic approach and conservation laws. Note that this derivation of the coupling conditions will change for another system of equations like the nonlinear shallow water equations. Finally we compute the evolution of kinks and breathers in a particular graph.

The article is organized as follows. In the following Section~\ref{sec:sg} we present some basic facts on the \acs{sg} equation and justify the interface conditions. In Section~\ref{sec:discr}, we introduce our detailed methodology to solve \textsc{Hamiltonian} evolution equations on networks. We apply it to the \acs{sg} equation. Numerical solutions for kinks and breathers on a given graph are shown in Section~\ref{sec:num} and we discuss these results in Section~\ref{sec:concl}.


\section{Continuous sine--Gordon equation}
\label{sec:sg}

Consider the real space-time coordinates $(x,\,t)\ \in\ \R\times\R^{\,+}\,$. Then, the most common version of the \acf{sg} equation reads \cite{Scott2003}
\begin{equation}\label{eq:sg}
  u_{\,t\,t}\ -\ u_{\,x\,x}\ +\ \sin u \ =\ 0\,,
\end{equation}
where the subscripts $(\cdot)_{t}\,$, $(\cdot)_{x}$ denote the derivatives with respect to the time $t$ and space $x$ coordinates. In order to obtain a well-posed boundary value problem, equation \eqref{eq:sg} is completed by periodic or homogeneous \textsc{Neumann} boundary conditions. The linear part $\Box^{\,2} u\ \eqdef\ u_{\,t\,t}\ -\ u_{\,x\,x}$ is the \textsc{D'Alembertian} or \textsc{Laplacian} in \textsc{Minkowski} space $\M^2\,$. The \acs{sg} equation is known to be \textsc{Lorentz} invariant and an integrable PDE \cite{Faddeev1987}.


\subsection{Variational structure}

The \acs{sg} equation can be derived as the \textsc{Euler}--\textsc{Lagrange} equation of the following \textsc{Lagrangian} density\footnote{One can notice that this \textsc{Lagrangian} is classical since it can be seen as the kinetic minus potential energies. See, for example \cite{Scott2004}.}
\begin{equation}\label{eq:ld}
  \L_{\,\mathrm{sG}}\ \eqdef\ \half\,\bigl(u_{\,t}^{\,2}\ -\ u_{\,x}^{\,2}\bigr)\ -\ 1\ +\ \cos u\,.
\end{equation}
Moreover, the \acs{sg} equation possesses also the \textsc{Hamiltonian} formulation
\begin{equation}\label{eq:ham}
  \z_{\,t}\ =\ \J\vd_\z\,\H\,, \qquad \J\ \eqdef\ 
  \begin{pmatrix}
     0 & 1 \\
    -1 & 0
   \end{pmatrix}\,,
\end{equation}
$\z\ \eqdef\ (u,\,v)\,$, $\vd_\z\ \eqdef\ (\delta_u,\, \delta_v)$ is the variational gradient, $\J$ is the symplectic operator and, finally, the \textsc{Hamiltonian} functional $\H$ is defined as
\begin{equation*}
  \H \{\,\z\,\}\ \eqdef\ \int_{-\infty}^{+\infty}\Bigl[\,\half\, v^{\,2}\ +\ \half\, u_{\,x}^{\,2}\ +\ 1\ -\ \cos u\,\Bigr]\, \ud x\,.
\end{equation*}
It can be obtained by a \textsc{Legendre} transform. Equations \eqref{eq:ham} can be rewritten component-wise form for the sake of clarity
\begin{eqnarray*}
  u_{\,t}\ &=&\ \frac{\delta\H}{\delta v}\ =\ v\,, \\
  v_{\,t}\ &=&\ -\frac{\delta\H}{\delta u}\ =\ u_{\,x\,x}\ -\ \sin u\,.
\end{eqnarray*}
Consequently, the \acs{sg} equation is a \textsc{Hamiltonian} system with phase space $(u,\,v)$ and the symplectic form
\begin{equation}\label{eq:sympl}
  \omega\ \eqdef\ \int_{-\infty}^{+\infty} \ud u \wedge \ud v\, \ud x\,.
\end{equation}
The multi-symplectic structure of the \acs{sg} equation is discussed in \cite{Islas2003}.

The \acs{sg} equation has an infinite number of conserved quantities \cite{Ablowitz1973}. It is therefore an integrable infinite dimensional \textsc{Hamiltonian} system. Among the conserved quantities two are particularly important, the \textsc{Hamiltonian} and the momentum. The \textsc{Hamiltonian} $\H\,\{u,\,v\} \equiv \E\{u\}$ has the sense of the physical energy
\begin{equation*}
  \E\{\,u\,\}\ \eqdef\ \int_{-\infty}^{+\infty}\Bigl[\,\half\, u_{\,t}^{\,2}\ +\ \half\, u_{\,x}^{\,2}\ +\ 1\ -\ \cos u\,\Bigr]\, \ud x\,.
\end{equation*}

\begin{remark}
There is another important functional conserved for the equation \eqref{eq:sg} which can be associated to the total momentum
\begin{equation*}
  \Mo\{\,\z\,\}\ \eqdef\ \int_{-\infty}^{+\infty} u_{\,t}\, u_{\,x}\, \ud x\,.
\end{equation*}
The conservation of $\Mo\,\{\,\z\,\}$ can be readily checked by computing $\od{\Mo}{t}\,$.
\end{remark}


\subsection{Exact solutions}

The \acs{sg} equation has constant solutions\footnote{After substituting a constant solution ansatz $u\,(x,\,t)\ \equiv\ C$ into \acs{sg} equation \eqref{eq:sg} we obtain that necessarily $\sin\,C\ =\ 0\,$. Thus, $C\ =\ l\,\pi$ with $l\ \in\ \Z\,$. However, only even values of $l\ =\ 2\,k\,$, $k\ \in\ \Z$ correspond to the minima of the potential energy. So, we keep only this sub-family of constant solutions.}
\begin{equation}\label{eq:stat}
  u\,(x,\,t)\ \equiv\ 2\, k\, \pi\,,
\end{equation}
where $k\ \in\ \Z$ is an integer. These have zero energy and are ground states. A first non trivial (\ie non-constant) solution is the \emph{kink} \cite{Scott2004}:
\begin{equation}\label{eq:kink0}
  u\,(x,\,t)\ =\ 4\arctan \ue^{\,\gamma\,(x\ -\ x_{\,0}\ -\ c\,t)}\,, \qquad x_{\,0}\ \in\ \R\,,
\end{equation}
where $c\ \in\ [0,\,1)$ is the kink celerity and $\gamma$ is the so-called \textsc{Lorentz} factor
\begin{equation}\label{eq:kink}
  \gamma^{\,2}\ \eqdef\ \frac{1}{1\ -\ c^{\,2}}\,.
\end{equation}
The energy of the kink can be computed analytically, $\E\ =\ 8\,\gamma\,$. In our scaling the speed of light is equal to $1\,$. The kink solution realizes a smooth transition between the two ground states $0\ \leadsto\ 2\,\pi\,$. More generally, kinks link in phase space two neighbouring states $2\,\pi\, k\ \leadsto\ 2\,\pi\,(k\ \pm\ 1)\,$.

There is also another type of exact solutions to the \acs{sg} equation --- the \emph{breathers} \cite{Ablowitz1973}. These solutions are localized and oscillate in space and in time; their analytical expression is given by \cite{Wattis1996}:
\begin{equation*}
  u\,(x,\,t)\ =\ 4\arctan\,\biggl\{\,\tan\mu\;\frac{\cos\bigl(\omega\,\gamma\,(t\ +\ x_{\,0}\,c\ -\ x\,c)\bigr)}{\cosh\bigl(\sin\mu\cdot\gamma\,(x\ -\ x_{\,0}\ -\ c\,t)\bigr)}\,\biggr\}\,, \qquad x_0\ \in\ \R\,,
\end{equation*}
where the parameter $\mu$ is defined through the relation $\cos\mu\ \equiv\ \omega\,$. The energy of a breather depends both on its speed and frequency as:
\begin{equation*}
  \E\ =\ 16\,\gamma\,\sqrt{1\ -\ \omega^2}\,.
\end{equation*}


\subsection{Coupling conditions at the junctions}

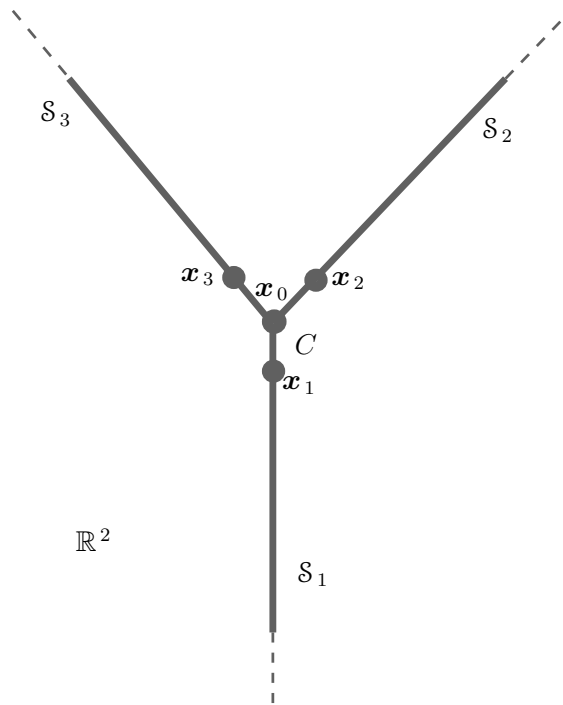
\begin{figure}
  \centering
  \scalebox{0.9} 
  {
  \begin{pspicture}(0,-5.12)(8.962812,5.12)
  \definecolor{color282}{rgb}{0.3764705882352941,0.3764705882352941,0.3764705882352941}
  \psline[linewidth=0.1cm,linecolor=color282](4.3809376,0.48)(1.4009376,4.1)
  \psline[linewidth=0.1cm,linecolor=color282](4.3809376,0.5)(7.7809377,4.1)
  \psline[linewidth=0.1cm,linecolor=color282](4.3809376,0.48)(4.3809376,-4.06)
  \psline[linewidth=0.04cm,linecolor=color282,linestyle=dashed,dash=0.16cm 0.16cm](1.3609375,4.16)(0.5809375,5.1)
  \psline[linewidth=0.04cm,linecolor=color282,linestyle=dashed,dash=0.16cm 0.16cm](7.7809377,4.1)(8.580937,4.94)
  \psline[linewidth=0.04cm,linecolor=color282,linestyle=dashed,dash=0.16cm 0.16cm](4.3809376,-4.06)(4.3809376,-5.1)
  \pscircle[linewidth=0.146,linecolor=color282,dimen=outer](4.4009376,0.52){0.18}
  \usefont{T1}{ptm}{m}{n}
  \rput(4.862344,0.185){$C$}
  \usefont{T1}{ptm}{m}{n}
  \rput(7.6723437,3.345){$\mathcal{S}_{\,2}$}
  \usefont{T1}{ptm}{m}{n}
  \rput(1.2123437,3.585){$\mathcal{S}_{\,3}$}
  \usefont{T1}{ptm}{m}{n}
  \rput(4.972344,-3.195){$\mathcal{S}_{\,1}$}
  \usefont{T1}{ptm}{m}{n}
  \rput(1.7623438,-2.695){$\mathbb{R}^{\,2}$}
  \pscircle[linewidth=0.146,linecolor=color282,dimen=outer](5.0109377,1.13){0.17}
  \pscircle[linewidth=0.146,linecolor=color282,dimen=outer](3.8109374,1.17){0.17}
  \pscircle[linewidth=0.146,linecolor=color282,dimen=outer](4.3909373,-0.21){0.17}
  \usefont{T1}{ptm}{m}{n}
  \rput(5.4823437,1.125){$\boldsymbol{x}_{\,2}$}
  \usefont{T1}{ptm}{m}{n}
  \rput(3.2823439,1.185){$\boldsymbol{x}_{\,3}$}
  \usefont{T1}{ptm}{m}{n}
  \rput(4.762344,-0.415){$\boldsymbol{x}_{\,1}$}
  \usefont{T1}{ptm}{m}{n}
  \rput(4.362344,0.96){$\boldsymbol{x}_{\,0}$}
  \end{pspicture} 
  }
  \caption{\small\em Sketch of a $Y-$junction composed of three strings $\S_{\,1,\,2,\,3}$ with the center located at the point $C\,$.}
  \label{fig:sketch1}
\end{figure}

We now consider that the \acs{sg} equation is defined on each branch $E$ of an oriented network $G\ =\ (V,\,E)$ where $V$ is a set of vertices and $E$ the set of branches. We label the branches $m\ =\ 1,\,\ldots,\, \abs{E}\,$. To fix the notations, consider a $Y-$junction composed of three semi-infinite rays $\S_{\,1,\,2,\,3}$ embedded into the \textsc{Euclidean} space $\R^{\,2}\,$:
\begin{equation*}
  \Y\ \eqdef\ \Bigl\{\x\,\in\,\R^{\,2}\,:\, \exists\, i\,\in\,\{1,\,2,\,3\} \mbox{ such that } \x\,\in\,\S_{\,i}\Bigr\}\,.
\end{equation*}
The intersection of all three strings is located at the unique point $C\ \in\ \R^{\,2}$ defined as (see Figure~\ref{fig:sketch1} for the illustration):
\begin{equation*}
  C\ \eqdef\ \S_{\,1}\ \cap\ \S_{\,2}\ \cap\ \S_{\,3}\,.
\end{equation*}
Henceforth, each ray $\S_{\,i}$ starts at the junction point $C$ and continues, for the sake of simplicity, indefinitely in the prescribed direction\footnote{Of course, this idealization is adopted only for the problem formulation. In our numerical simulations we assume the branches $\S_{\,i}$ to be finite with length $\ell_{\,i}\,$.}. A simple topological argument can be applied to show that the set $\Y$ is not homeomorphic to any \textsc{Euclidean} space. Indeed, let us remove virtually the junction point $C$ from this set $\Y\,$. It will be decomposed in three disjoint components. Clearly, any \textsc{Euclidean} space $\R^{\,n}\,$, $n\ \geq\ 1$ does not have such a point with the same property.

A first natural condition to be satisfied at the junction point $C$ is the continuity of the solution
\begin{equation}\label{eq:cont}
  \lim\limits_{\x\,\to\,C,\; \x\,\in\,\S_{\,1}} u\,(x,\,t)\ =\ \lim\limits_{\x\,\to\,C,\; \x\,\in\,\S_{\,2}} u\,(x,\,t)\ =\ \lim\limits_{\x\,\to\,C,\; \x\,\in\,\S_{\,3}} u\,(x,\,t)\,.
\end{equation}
Moreover, this condition has to be completed by the ``\emph{charge}'' conservation property adopted in previous studies of the \textsc{Klein}--\textsc{Gordon} \cite{Mehmeti2003, Bibikov2009} and \acs{sg} \cite{Caputo2014} equations:
\begin{equation}\label{eq:kirch}
  \left.\partial_{\,1} u \right\vert_{\x\,\to\,C,\; \x\,\in\,\S_{\,1}}\ +\ \left.\partial_{\,2} u \right\vert_{\x\,\to\,C,\; \x\,\in\,\S_{\,2}}\ +\ \left.\partial_{\,3} u \right\vert_{\x\,\to\,C,\; \x\,\in\,\S_{\,3}}\ =\ 0\,,
\end{equation}
where $\partial_{\,i}$ denotes the first spatial derivative along the branch $\S_{\,i}\,$, $i\ =\ 1,\,2,\,3\,$. Condition \eqref{eq:kirch} is a continuous analogue of the celebrated \textsc{Kirchhoff}'s circuit law. We will justify these conditions in the next two sections using an homothetic argument and conservation laws respectively.

\begin{figure}
  \centering
  \includegraphics[width=0.75\textwidth]{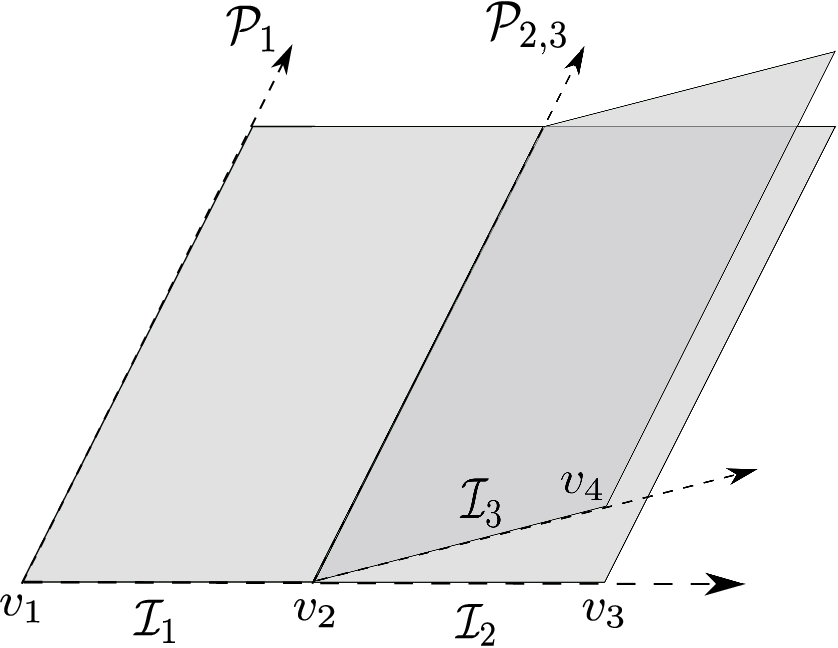}
  \caption{\small\em A schematic representation of the foliation of the phase spaces at a junction point.}
  \label{fig:phase}
\end{figure}

Before justifying the coupling conditions, let us examine the structure of the phase space associated to the \acs{sg} dynamics on our $Y-$junction, \ie the center of Figure~\ref{fig:merc}. The augmented phase space of the \acs{sg} equation restricted to a branch (say $e_{\,1}$) consists in $\P \times \In\,$, where $\P\ \eqdef\ (u_{\,t\,t},\, u_{\,t},\, u,\, u_{\,x},\, u_{\,x\,x})$ is the usual configuration space and $\In$ is the spatial extent of a branch ($\In\ =\ [\,v_{\,1},\, v_{\,2}\,]$ in this particular case). We have to augment the configuration space $\P$ by $\In$ since the joints are realized on the boundaries of the interval $\In\,$. This portion of the phase space is schematically represented on Figure~\ref{fig:phase} for the $Y-$junction; in Geometry these structures are called \emph{foliations} \cite{Haefliger1970}. This is only the local structure of the global phase space; to represent the global phase space associated to a network, the borders of the phase spaces of the individual branches, the leaves on Figure~\ref{fig:phase}, would have to be glued together in accordance with the scheme prescribed by the graph $G\,$.

\subsubsection{Back-to-Manifold: a homothetic approach}
\label{sec:man}

\begin{figure}
  \centering
  \scalebox{0.85} 
  {
  \begin{pspicture}(0,-5.426683)(10.822812,5.426269)
  \definecolor{color183}{rgb}{0.6745098039215687,0.0,0.0}
  \definecolor{color1867b}{rgb}{0.6745098039215687,0.6745098039215687,0.6745098039215687}
  \definecolor{color1868}{rgb}{0.6352941176470588,0.6352941176470588,0.6352941176470588}
  \definecolor{color1882}{rgb}{0.00392156862745098,0.00392156862745098,0.00392156862745098}
  \definecolor{color1918}{rgb}{0.6745098039215687,0.00784313725490196,0.00784313725490196}
  \psframe[linewidth=0.002,linecolor=color1867b,dimen=outer,fillstyle=solid,fillcolor=color1867b](6.2209377,0.19331692)(4.4209375,-5.426683)
  \rput{-45.0}(0.23230292,6.167463){\psframe[linewidth=0.002,linecolor=color1868,linestyle=dotted,dotsep=0.16cm,dimen=outer,fillstyle=solid,fillcolor=color1867b](8.4609375,5.613317)(6.6609373,-0.006683079)}
  \rput{45.0}(2.8846302,-1.3574793){\psframe[linewidth=0.002,linecolor=color1868,linestyle=dotted,dotsep=0.16cm,dimen=outer,fillstyle=solid,fillcolor=color1867b](3.9809375,5.613317)(2.1809375,-0.006683079)}
  \psline[linewidth=0.002,linecolor=color1867b,fillstyle=solid,fillcolor=color1867b](6.2209377,0.17331693)(4.4209375,0.17331693)(5.3209376,1.153317)(6.2209377,0.19331692)
  \usefont{T1}{ptm}{m}{n}
  \rput(9.532344,4.0783167){$\mathcal{S}_{\,2}$}
  \usefont{T1}{ptm}{m}{n}
  \rput(4.7723436,-5.061683){$\mathcal{S}_{\,1}$}
  \usefont{T1}{ptm}{m}{n}
  \rput(1.2123437,4.118317){$\mathcal{S}_{\,3}$}
  \psline[linewidth=0.02cm,linecolor=color1882,linestyle=dashed,dash=0.16cm 0.16cm,arrowsize=0.05291667cm 2.0,arrowlength=1.4,arrowinset=0.4]{<->}(7.0009375,3.493317)(8.240937,2.2533169)
  \usefont{T1}{ptm}{m}{n}
  \rput(7.8923435,3.078317){$\delta$}
  \pscircle[linewidth=0.02,linecolor=color1882,linestyle=dashed,dash=0.16cm 0.16cm,dimen=outer](5.2809377,0.8933169){2.0}
  \pscircle[linewidth=0.022,linecolor=color1882,dimen=outer,fillstyle=solid,fillcolor=color1882](5.3409376,0.75331694){0.08}
  \usefont{T1}{ptm}{m}{n}
  \rput(5.9423437,1.0183169){$C$}
  \usefont{T1}{ptm}{m}{n}
  \rput(2.2223437,-4.041683){$\mathbb{R}^{\,2}$}
  \usefont{T1}{ptm}{m}{n}
  \rput(7.4023438,-0.14168307){$\Omega_{\,\delta}$}
  \psline[linewidth=0.02cm,linecolor=color1882,linestyle=dashed,dash=0.16cm 0.16cm,arrowsize=0.05291667cm 2.0,arrowlength=1.4,arrowinset=0.4]{<->}(2.3609376,2.2733169)(3.5609374,3.5733168)
  \usefont{T1}{ptm}{m}{n}
  \rput(2.7123437,3.1383169){$\delta$}
  \psline[linewidth=0.02cm,linecolor=color1882,linestyle=dashed,dash=0.16cm 0.16cm,arrowsize=0.05291667cm 2.0,arrowlength=1.4,arrowinset=0.4]{<->}(4.4409375,-1.8666831)(6.2409377,-1.846683)
  \usefont{T1}{ptm}{m}{n}
  \rput(5.3323436,-2.141683){$\delta$}
  \psline[linewidth=0.02cm,linecolor=color1882,linestyle=dashed,dash=0.16cm 0.16cm,arrowsize=0.05291667cm 2.0,arrowlength=1.4,arrowinset=0.4]{->}(5.2609377,0.8333169)(3.7009375,2.113317)
  \usefont{T1}{ptm}{m}{n}
  \rput(4.762344,1.678317){$2\,\delta$}
  \usefont{T1}{ptm}{m}{n}
  \rput(5.242344,-3.421683){$\mathcal{Y}_{\,\delta}$}
  \rput{-45.0}(0.9004505,4.0028687){\psarc[linewidth=0.06,linecolor=color183](5.282115,0.9144943){1.9798989}{53.020363}{108.954254}}
  \rput{45.0}(2.214508,-3.3712766){\psarc[linewidth=0.06,linecolor=color1918](5.176745,0.98750955){1.8596908}{67.62315}{126.45429}}
  \rput{180.0}(10.521875,1.7466339){\psarc[linewidth=0.06,linecolor=color183](5.2609377,0.87331694){1.98}{65.897766}{119.43849}}
  \usefont{T1}{ptm}{m}{n}
  \rput(5.2923436,-1.4216831){\color{color183}$\ell_{\,1}$}
  \usefont{T1}{ptm}{m}{n}
  \rput(7.132344,2.298317){\color{color183}$\ell_{\,2}$}
  \usefont{T1}{ptm}{m}{n}
  \rput(3.4523437,2.3783169){\color{color183}$\ell_{\,3}$}
  \end{pspicture} 
  }
  \caption{\small\em The $Y-$junction from the previous Figure~\ref{fig:sketch1} converted into a manifold $\Y_{\,\delta}$ by extending all the branches to some finite thickness $\delta\ >\ 0\,$.}
  \label{fig:sketch2}
\end{figure}
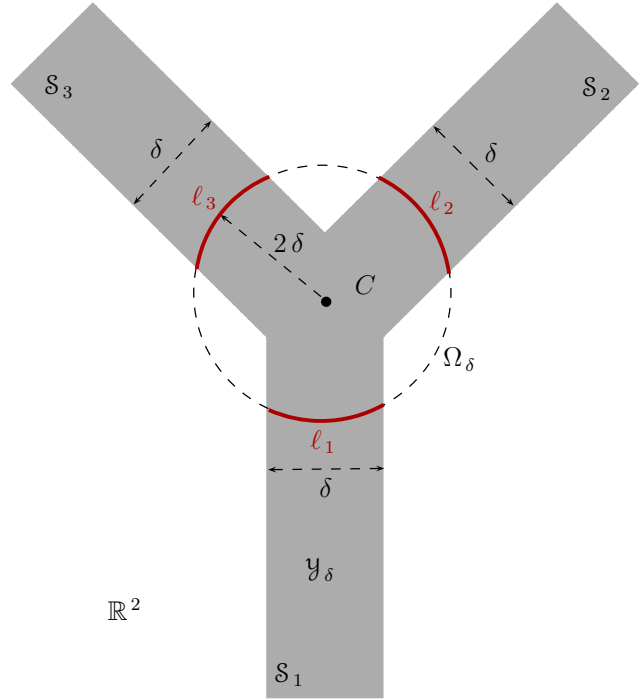

Condition \eqref{eq:kirch} can be justified by converting (`\emph{inflating}') the $Y-$junction domain into a manifold $\Y_\delta$ of small thickness $\delta\ >\ 0\,$. Thus, $\Y_\delta$ becomes a tubular neighbourhood\footnote{This tubular neighbourhood should be considered as a surface.} of our network. Moreover, the two-dimensional version of the \acs{sg} equation becomes
\begin{equation}\label{eq:multi}
  u_{\,t\,t}\ -\ \grad^2\,u\ +\ \sin u \ =\ 0\,.
\end{equation}
The elements necessary to the proof are shown on Figure~\ref{fig:sketch2}. On the boundary $\partial\Y_{\delta}$ we impose the homogeneous \textsc{Neumann} condition \cite{Gulevich2006, Caputo2014}:
\begin{equation}\label{eq:neuman}
  \left.\partial_n\, u\,\right|_{\x\ \in\ \partial\,\Y_\delta}\ =\ 0\,, \qquad
  \partial_{\,n}\, u\ \eqdef\ \grad u\scal \n\,,
\end{equation}

To justify \eqref{eq:kirch} we integrate the 2D version of the \acs{sg} equation \ref{eq:multi} over the domain $\Omega_\delta\ \cap\ \Y_\delta$\footnote{The domain $\Omega_\delta\ \eqdef\ \D_{\,2\,\delta}\,(C)$ is the disc of radius $2\,\delta$ centered at the junction point $C$ depicted on Figure~\ref{fig:sketch2}.} and use the \textsc{Stokes} theorem for the divergence (\textsc{Laplacian}) term:
\begin{equation*}
  \iint\limits_{\Omega_{\,\delta}\ \cap\ \Y_{\,\delta}}\Bigl[u_{\,t\,t}\ -\ \grad^2 u\ +\ \sin u\Bigr]\;\ud\x\ =\ \underbrace{\iint\limits_{\Omega_{\,\delta}\ \cap\ \Y_{\,\delta}}\Bigl[u_{\,t\,t}\ +\ \sin u\Bigr]\,\ud\x}_{(i)} + \underbrace{\int\limits_{\partial(\Omega_{\,\delta}\ \cap\ \Y_{\,\delta})}\partial_{\,n}\, u\;\ud s}_{(ii)}\ =\ 0\,.
\end{equation*}
Assuming the solution $u\,(\x,\,t)$ to be smooth and bounded in $\Omega_{\,\delta}\,$, the first integral $(i)$ scales as $\O\,(\delta^2)\,$. Taking into account the boundary condition \eqref{eq:neuman}, the second integral $(ii)$ becomes just the sum of three line integrals over $\ell_i$ in the interior of the domain $\Y_{\,\delta}$ (represented in red on Figure~\ref{fig:sketch2}). Thus the sum of the two integrals reduces to
\begin{equation*}
  \O\,(\delta^{\,2})\ +\ \sum_{i\,=\,1}^3\underbrace{\int_{\ell_i}\partial_{\,n} u\;\ud s}_{\O\,(\delta)}\ =\ 0\,,
\end{equation*}
where under the same assumptions on the solution $u\,(x,\,t)$, the integrals in the second sum scale as $\O(\delta)\,$. Thus, dividing the identity by $\delta$ and taking the limit $\delta\ \to\ 0\,$, we obtain the desired result \eqref{eq:kirch}. An alternative derivation based on the variational structure can be found in Appendix~\ref{app:cons}.

\subsubsection{Conservation laws approach}
\label{sec:law}

\vspace*{0.5em}
\paragraph*{Energy.}

The main conserved quantity, the energy can be used to justify the interface conditions \eqref{eq:kirch}. The energy of a network is
\begin{equation*}
  \E\ =\ \sum_{i\,=\,1}^{m}\; \int_{a_{\,i}}^{b_{\,i}} \bigl[\,\half\bigl(u_{\,t}^{\,2}\ +\ u_{\,x}^{\,2}\bigr)\ +\ 1\ -\ \cos\,u\,\bigr]\;\ud x\,, \qquad \ell_{\,i}\ \eqdef\ b_{\,i}\ -\ a_{\,i}\,,
\end{equation*}
where the sum is taken over all $m$ branches of the network and the branch $i$ starts from the node $a_{\,i}$ and finishes at the node $b_{\,i}\,$. Taking the time derivative of $\E$ we get
\begin{equation*}
  \od{\E}{t}\ =\ \sum_{i\,=\,1}^{m}\;\bigl[\,u_{\,t}\,u_{\,x}\,\bigr]_{\,a_{\,i}}^{\,b_{\,i}}\,.
\end{equation*}
We can assume that the nodes $a_{\,i}$ and $b_{\,i}$ are uncoupled. This is natural since information can only travel from $a_{\,i}$ to $b_{\,i}$ at a finite speed. Therefore, to satisfy $\od{\E}{t}\ =\ 0$ we need that
\begin{equation*}
  \sum_{i}\;u_{\,t}\,u_{\,x}\bigr\vert_{\,a_{\,i}}\ =\ \sum_{i}\;u_{\,t}\,u_{\,x}\bigr\vert_{\,b_{\,i}}\ \equiv\ 0\,.
\end{equation*}
For a $Y-$junction, this can be satisfied if $u_{\,t}$ is continuous, \ie $u_{\,1\,t}\ =\ u_{\,2\,t}\ =\ u_{\,3\,t}$ and $-{u_{\,1}}_{\,x}\ +\ {u_{\,2}}_{\,x}\ +\ {u_{\,3}}_{\,x}\ =\ 0\,$, which is precisely the \textsc{Kirchoff} law \eqref{eq:kirch}.

These coupling conditions \eqref{eq:cont} and \eqref{eq:kirch} are natural in terms of the electrodynamics of \textsc{Josephson} junctions where the variable $u$ is the phase difference between the two superconductors and where $u_{\,x}$ is the surface current in the junction. The conditions \eqref{eq:cont} and \eqref{eq:kirch} state that the phase is continuous and that the currents satisfy \textsc{Kirchoff}'s law.

\vspace*{0.5em}
\paragraph*{Momentum.}

One could try to impose similarly the conservation of the total momentum of the network:
\begin{equation*}
  \Mo\ =\ \sum_{i\,=\,1}^{m}\;\int_{\,a_{\,i}}^{\,b_{\,i}}\; u_{\,t}\, u_{\,x}\; \ud\,x\,.
\end{equation*}
We can compute $\od{\Mo}{t}\,$, impose momentum conservation condition $\od{\Mo}{t}\ =\ 0$ and follow the same procedure as above and obtain
\begin{equation*}
  \sum_{i=1}^m\;\bigl[\,\half\bigl(u_{\,t}^{\,2}\ +\ u_{\,x}^{\,2}\bigr)\ -\ 1\ +\ \cos\u\,\bigr]\bigr\vert_{\,a_{\,i}}\ =\ \sum_{i}^m\;\bigl[\,\half\bigl(u_{\,t}^{\,2}\ +\ u_{\,x}^{\,2}\bigr)\ -\ 1\ +\ \cos\u\,\bigr]\bigr\vert_{\,b_{\,i}}\,.
\end{equation*}
We then get for the $Y-$junction, taking into account \eqref{eq:cont} and \eqref{eq:kirch}: 
\begin{equation*}
  -\,u_{\,1\,x}^{\,2}\ +\ u_{\,2\,x}^{\,2}\ +\ u_{\,3\,x}^{\,2}\ =\ 0\,,
\end{equation*}
which cannot be satisfied in general. This problem exists also for the shallow water equations in a junction \cite{Caputo2016}. We then see that on a network we loose a number of conserved quantities.

In the sequel of the paper we require continuity \eqref{eq:cont} and `\emph{charge}' conservation \eqref{eq:kirch} at the junction points. We also remark the following:
\begin{itemize}
  \item This approach can be generalized to any general nonlinearity in equation \eqref{eq:sg}, not only $f\,(u)\ =\ \sin u\,$:
  \begin{equation*}
    u_{\,t\,t}\ -\ u_{\,x\,x}\ +\ f\,(u) \ =\ 0\,.
  \end{equation*}
  \item This approach can also be applied to non-\textsc{Hamiltonian} systems. There one should consider other conservation laws, see for example the nonlinear shallow water equations \cite{Caputo2016}.
  \item The compatibility conditions \eqref{eq:cont}, \eqref{eq:kirch} at the graph vertices can be straightforwardly generalized to situations where any (finite) number of strings meet at one junction point. Another generalization consists in assigning different weights $\omega_{\,i}\,$, $i\ =\ 1,\,2,\,3,\,\ldots$ to incident edges. They can be interpreted as widths of channels, for example. The derivation of coupling conditions in this case can be found \eg in \cite{Caputo2014}.
  \item It goes without saying that the initial condition $u\,(\x,\,t\,=\,0)\ =\ u_{\,0}\,(\x)$ on a graph $G$ should satisfy conditions \eqref{eq:cont} and \eqref{eq:kirch}.
\end{itemize}


\section{Numerical implementation : the discrete sine-Gordon equation}
\label{sec:discr}

Traditionally, the \acs{sg} equation was solved numerically among others using finite difference \cite{Furihata2001}, finite element \cite{Argyris1991, Caputo2014}, tension spline \cite{Rashidinia2011} and radial basis functions \cite{Dehghan2008, Ilati2015} methods. In order to propose a discrete version of the \acs{sg} equation we will follow the variational framework. Recall that the \acs{sg} equation is a \textsc{Hamiltonian} PDE. A natural way to convert it into a discrete dynamical system is to employ a symplectic discretization \cite{Leimkuhler2004}. The work-flow is determined by the method of lines:
\begin{itemize}
  \item Discretize the \textsc{Hamiltonian} functional $\H\,[\,\z\,]$ in space on a lattice to obtain a system of coupled \textsc{Hamiltonian} ODEs
  \item Discretization in time the system of \textsc{Hamiltonian} ODEs using a symplectic scheme.
\end{itemize}
This programme will be realized below by following the main lines of \cite{Leimkuhler2004}. Please, notice also the differences between symplectic and variational integrators \cite{Lew2003, Marsden1998}.

Consider a uniform lattice $\{x_{\,j}\ =\ k\Delta x\, \vert\, j\,=\,1,\,\ldots,\,n\}\,$, $\Delta x\ >\ 0\,$. The values of the \acs{sg} solution $u(x,\,t)$ at lattice points will be denoted by $u_{\,j}\ \approx\ u\,(x_{\,j})\,$. For the moment we will consider only interior nodes. The junction points (end points of the lattice) will be discussed below. After the discretization, the phase space becomes finite dimensional, since $\{\z_{\,j}\}_{j\,=\,1}^{\,n}\ =\ \{(u_{\,j},\, v_{\,j})\;\in\;\R^{\,2}\}_{j\,=\,1}^{\,n}\ \in\ \R^{\,2\,n}\,$. The discrete symplectic form on this space becomes
\begin{equation*}
  \omega_{\,n}\ =\ \sum_{j\,=\,1}^{n}\ud u_{\,j}\ \wedge\ \ud v_{\,j}\, \Delta x\,,
\end{equation*}
which is a straightforward discretization of \eqref{eq:sympl}. The \textsc{Hamiltonian} functional will be approximated with the \emph{rectangular rule} as the following sum
\begin{equation*}
  \H_{\,n}\,[\{u_{\,j},\,v_{\,j}\}]\ =\ \sum_{j\,=\,1}^{n}\biggl[\half\, v_{\,j}^{\,2}\ +\ \half\,\Bigl(\frac{u_{\,j}\ -\ u_{\,j-1}}{\Delta x}\Bigr)^2\ +\ 1\ -\ \cos u_{\,j} \biggr]\,\Delta x\,.
\end{equation*}
The system of \textsc{Hamiltonian} ODEs follows automatically
\begin{equation*}
  \od{\z_{\,j}}{t}\ =\ \J_{\,n}\cdot\grad_{\z_{\,j}}\,\H_{\,n}\,[\,\z\,]\,, \qquad
  \J_{\,n}\ =\ \begin{pmatrix}
           \boldsymbol{0} & \I_{\,n} \\
           -\I_{\,n} & \boldsymbol{0}
         \end{pmatrix}, \qquad i\,=\,1,\ldots,n\,.
\end{equation*}
After computing the derivatives, the last semi-discrete system becomes
\begin{eqnarray}\label{eq:semid1}
  \od{u_{\,j}}{t} &=& v_{\,j}\,, \\
  \od{v_{\,j}}{t} &=& \frac{w_{\,j+1}\ -\ w_{\,j}}{\Delta x}\ -\ \sin u_{\,j}\,, \label{eq:semid2}
\end{eqnarray}
where $w_{\,j+1}\,(t)\ \eqdef\ \dfrac{u_{\,j+1}\ -\ u_{\,j}}{\Delta x}$ and $w_{\,j}\,(t)\ \eqdef\ \dfrac{u_{\,j}\ -\ u_{\,j-1}}{\Delta x}\,$. It can be shown \cite{Leimkuhler2004} that the semi-discrete scheme \eqref{eq:semid1}, \eqref{eq:semid2} satisfies a local energy conservation law
\begin{equation}\label{eq:ed}
  \od{}{t}\;\Bigl[\,\half\, v_{\,j}^{\,2}\ +\ \half\, w_{\,j}^{\,2}\ +\ 1\ -\ \cos u_{\,j}\,\Bigr]\ +\ \frac{F_{\,j+\frac12}\ -\ F_{\,j-\frac12}}{\Delta x}\ =\ 0\,,
\end{equation}
where the quantity in brackets $\E_{\,j}\ \eqdef\ \half\, v_{\,j}^{\,2}\ +\ \half\, w_{\,j}^{\,2}\ +\ 1\ -\ \cos u_{\,j}$ is the (semi-)discrete energy and $F_{\,j+\frac12}\ \eqdef\ -\,v_{\,j}\,w_{\,j+1}\,$, $F_{\,j-\frac12}\ \eqdef\ -\,v_{\,j-1}\,w_{\,j}$ are the energy fluxes.

In order to obtain a fully discrete scheme, the system of ODEs \eqref{eq:semid1}, \eqref{eq:semid2} can be discretized in time with a \emph{symplectic Euler} method, for example:
\begin{eqnarray*}
  u_{\,j}^{\,m+1}\ &=&\ u_{\,j}^{\,m}\ +\ \Delta t\, v_{\,j}^{\,m+1}\,, \\
  v_{\,j}^{\,m+1}\ &=&\ v_{\,j}^{\,m}\ +\ \Delta t\, \biggl[\frac{w_{\,j+1}^{\,m}\ -\ w_{\,j}^{\,m}}{\Delta x}\ -\ \sin u_{\,j}^{\,m}\biggr]\,,
\end{eqnarray*}
where $\Delta t\ >\ 0$ and $u_{\,j}^{\,m}\ \eqdef\ u\,(x_{\,j},\, t_{\,m})\,$, $t_{\,m}\ \eqdef\ m\,\Delta t\,$, $m\ =\ 1,\,2,\,\ldots$ After eliminating $v_{\,j}^{\,n}$ and $w_{\,j}^{\,n}$ from these equations we obtain the classical leap-frog scheme as a fully discrete analog of the \acs{sg} equation
\begin{equation*}
  \frac{u_{\,j}^{\,m+1}\ -\ 2u_{\,j}^{\,m}\ +\ u_{\,j}^{\,m-1}}{\Delta t^{\,2}}\ -\ \frac{u_{\,j+1}^{\,m}\ -\ 2\,u_{\,j}^{\,m}\ +\ u_{\,j-1}^{\,m}}{\Delta x^{\,2}}\ +\ \sin u_{\,j}^{\,m}\ =\ 0\,.
\end{equation*}
After simple algebraic manipulations we can obtain the following discrete dynamical system for the interior nodes of the lattice:
\begin{multline}\label{eq:dsg}
  u_{\,j}^{\,m+1}\ =\ 2\,u_{\,j}^{\,m}\ -\ u_{\,j}^{\,m-1}\ +\\ \Bigl(\,\frac{\Delta t}{\Delta x}\,\Bigr)^2\underbrace{\Bigl[\,u_{\,j+1}^{\,m}\ -\ 2\,u_{\,j}^{\,m}\ +\ u_{\,j-1}^{\,m}\,\Bigr]}_{\displaystyle{\eqdef\ \L\,u^{\,m}}}\ -\ \Delta t^{\,2}\,\sin u_{\,j}^{\,m}\,, \quad m\ =\ 1,\, 2, \ldots\,,
\end{multline}
where for the sake of simplicity we introduced the linear operator $\L\,$, which represents the discretization of the classical \textsc{Laplace} operator. The treatment of nodes at junctions will be discussed below in Sections~\ref{sec:conds} and \ref{sec:algo}. We underline that our discretization being explicit is subject to a CFL-type condition on the time step \cite{Courant1928}. However, this restriction is of hyperbolic type, which is quite gentle, and we had no practical difficulties to satisfy it.

In general, one cannot expect to have the fully discrete energy conservation law similar to \eqref{eq:ed}, in contrast to the semi-discrete schemes. The reason is that a symplectic scheme aims to preserve the symplectic form and it does not guarantee anything about the \textsc{Hamiltonian}. However, the backward error analysis explains why, in general, the symplectic discretizations of PDEs show satisfactory energy conservation properties \cite{Moore2003}.


\subsection{Formulation on graphs}
\label{sec:graphs}

In this Section we will describe the assembling procedure of 1D lattices $\ell_{\,i}\ \eqdef\ \bigl\{\x_{\,j}\,\in\,[\a_{\,i}\, =\, \x_{\,1},\, \b_{\,i}\, =\, \x_{\,n_{\,i}}]\, \vert\, j\,=\,1,\,\ldots,\,n_{\,i}\bigr\}$ into a network whose mathematical description is usually given on the language of the graph theory. In the sequel we will denote by $\a_{\,i}\,$, $\b_{\,i}$ the starting and terminal points of the lattice $\ell_{\,i}$ respectively.

Consider a simple oriented network-shaped weighted graph $G\ :=\ (V,\,E)\,$. The finite set of vertices $V$ is basically the union of lattice initial and terminal nodes:
\begin{equation*}
  V\ \eqdef\ \{\v_j\}_{j\,=\,1}^{m}\ \equiv\ \bigcup\limits_{i\,=\,1}^{\abs{E}}\ \{\a_{\,i},\, \b_{\,i}\}\,.
\end{equation*}
The finite set of edges $E\ \subseteq\ \Bigl\{(\a,\, \b)\ \in\ V^{\,2}\ |\ \a\ \neq\ \b \Bigr\}\,$. Every edge $e_{\,i}\ \eqdef\ (\ell_{\,i},\, \omega_{\,i})$ consists of 1D lattice segments whose orientation is naturally determined by the enumeration of discrete lattice points (or equivalently the choice of the first and last points $\a_{\,i}$ and $\b_{\,i}$). The \emph{length} of the edge $\ell_{\,i}$ can be prescribed through its weight $\omega_{\,i}\,$:
\begin{equation*}
  \omega_{\,i}\ \eqdef\ \abs{\ell_{\,i}}\,\Delta x_{\,i}\ \equiv \ n_{\,i}\,, \qquad i\ =\ 1,\,\ldots,\, \abs{E}\,.
\end{equation*}
where $\abs{\ell_{\,i}}$ is the number of points in the lattice and $\Delta x_{\,i}$ is the spacing between two consecutive points.

\begin{figure}
  \centering
  \scalebox{0.85} 
  {
  \begin{pspicture}(0,-5.088125)(9.321875,5.088125)
  \definecolor{color1}{rgb}{0.3215686274509804,0.3215686274509804,0.3215686274509804}
  \definecolor{color162b}{rgb}{0.21176470588235294,0.21176470588235294,0.21176470588235294}
  \psline[linewidth=0.051999997cm,linecolor=color1,arrowsize=0.193cm 2.0,arrowlength=1.4,arrowinset=0.4]{>-}(4.48,-2.4503126)(4.48,-0.2703125)
  \psline[linewidth=0.06cm,linecolor=color1,arrowsize=0.193cm 2.0,arrowlength=1.4,arrowinset=0.4]{->}(4.48,-0.3103125)(6.366702,1.383194)
  \psline[linewidth=0.06cm,linecolor=color1,arrowsize=0.193cm 2.0,arrowlength=1.4,arrowinset=0.4]{->}(4.48,-0.2903125)(2.6952674,1.3510033)
  \psline[linewidth=0.06cm,linecolor=color1](6.2889,1.3204715)(8.019998,2.9094405)
  \psline[linewidth=0.06cm,linecolor=color1](2.78,1.2896875)(1.0597715,2.7896874)
  \psline[linewidth=0.06cm,linecolor=color1](4.48,-2.2703125)(4.5,-4.5103126)
  \pscircle[linewidth=0.06,linecolor=color1,dimen=outer](4.55,0.0){4.55}
  \pscircle[linewidth=0.06,linecolor=color162b,dimen=outer,fillstyle=solid,fillcolor=color162b](4.49,-0.2603125){0.13}
  \pscircle[linewidth=0.06,linecolor=color162b,dimen=outer,fillstyle=solid,fillcolor=color162b](8.03,2.8996875){0.13}
  \pscircle[linewidth=0.06,linecolor=color162b,dimen=outer,fillstyle=solid,fillcolor=color162b](4.51,-4.4803123){0.13}
  \pscircle[linewidth=0.06,linecolor=color162b,dimen=outer,fillstyle=solid,fillcolor=color162b](1.05,2.7996874){0.13}
  \psline[linewidth=0.06cm,linecolor=color1](4.28,4.5296874)(4.82,4.6096873)
  \psline[linewidth=0.06cm,linecolor=color1](4.28,4.5096874)(4.82,4.4296875)
  \psline[linewidth=0.06cm,linecolor=color1](1.1,-2.9103124)(0.98,-2.5903125)
  \psline[linewidth=0.06cm,linecolor=color1](1.12,-2.9103124)(0.82,-2.7303126)
  \psline[linewidth=0.06cm,linecolor=color1](8.34,-2.4503126)(8.08,-2.6703124)
  \psline[linewidth=0.06cm,linecolor=color1](8.34,-2.4503126)(8.24,-2.7903125)
  \usefont{T1}{ptm}{m}{n}
  \rput(4.5514064,-4.8653126){$v_{\,1}$}
  \usefont{T1}{ptm}{m}{n}
  \rput(5.0314064,-0.4653125){$v_{\,2}$}
  \usefont{T1}{ptm}{m}{n}
  \rput(8.511406,3.1946876){$v_{\,3}$}
  \usefont{T1}{ptm}{m}{n}
  \rput(0.83140624,3.0746875){$v_{\,4}$}
  \usefont{T1}{ptm}{m}{n}
  \rput(4.981406,-2.6853125){$e_{\,1}$}
  \usefont{T1}{ptm}{m}{n}
  \rput(8.841406,-2.8453126){$e_{\,6}$}
  \usefont{T1}{ptm}{m}{n}
  \rput(4.541406,4.8946877){$e_{\,4}$}
  \usefont{T1}{ptm}{m}{n}
  \rput(6.621406,0.9346875){$e_{\,2}$}
  \usefont{T1}{ptm}{m}{n}
  \rput(2.4814062,1.1546875){$e_{\,3}$}
  \usefont{T1}{ptm}{m}{n}
  \rput(0.52140623,-2.8453126){$e_{\,5}$}
  \end{pspicture} 
  }
  \caption{\small\em A sample graph used in our study for the sake of illustration.}
  \label{fig:merc}
\end{figure}
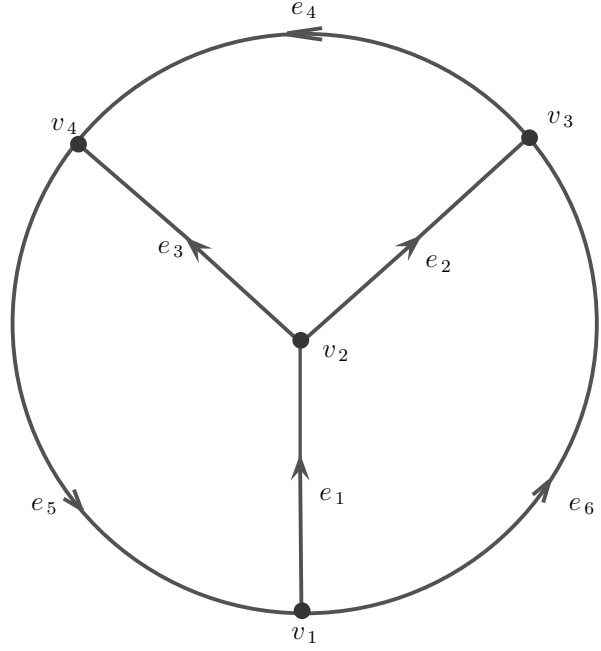

The internal organization of the network $G$ is traditionally given by graph-theoretical data structures such as the incidence and adjacency matrices \cite{Gould2012}. In the present study we will privilege the incidence matrix representation. By definition, the incidence matrix $A\ =\ \bigl(A_{\,i\,j}\bigr)_{n\times m}\ \in\ \Mat_{n_{\,e}\times m}\,(\Z)\,$, $n\ \eqdef\ \abs{V}\,$, $m\ \eqdef\ \abs{E}$ has the following elements:
\begin{equation*}
  A_{\,i\,j}\ =\ \begin{dcases}
  \ +1\,, & \mbox{ if the edge } e_{\,j} \mbox{ enters the vertex } v_{\,i}\,, \\
  \ -1\,, & \mbox{ if the edge } e_{\,j} \mbox{ leaves the vertex } v_{\,i}\,, \\
  \ 0\,, & \mbox { otherwise.}
  \end{dcases}
\end{equation*}
The edges are considered to be directional. For the sake of illustration let us consider the graph $G_{\,0}\ =\ (V,\,E)$ represented on Figure~\ref{fig:merc}. It is composed of four vertices $V\ =\ \{v_{\,1},\, v_{\,2},\, v_{\,3},\, v_{\,4}\}$ and six edges $E\ =\ \{e_{\,1},\,\ldots,\,e_{\,6}\}\,$. It is straightforward to check that the incidence matrix $\In_{\,0}$ of the graph $G_{\,0}$ is
\begin{equation*}
  \In_{\,0}\ =\ 
  \begin{pmatrix}
    -1 &  0 &  0 &  0 &  1 & -1 \\
     1 & -1 & -1 &  0 &  0 &  0 \\
     0 &  1 &  0 & -1 &  0 &  1 \\
     0 &  0 &  1 &  1 & -1 &  0
  \end{pmatrix}\,.
\end{equation*}
However, the most important information for us is the correspondence between the vertices $v_{\,i}$ with starting/terminal points of the lattices which compose the edges $e_{\,j}\,$. This correspondence is given as a list:
\begin{eqnarray*}
  v_{\,1}\ =\ \{\a_{\,1},\, \a_{\,6},\, \b_{\,5}\},\, & v_{\,3}\ =\ \{\a_{\,4},\, \b_{\,2},\, \b_{\,6}\}\,, \\
  v_{\,2}\ =\ \{\a_{\,2},\, \a_{\,3},\, \b_{\,1}\},\, & v_{\,4}\ =\ \{\a_{\,5},\, \b_{\,3},\, \b_{\,4}\}\,.
\end{eqnarray*}

\subsection{Conditions on junctions: the discrete case}
\label{sec:conds}

It is straightforward to obtain the discrete version of compatibility conditions \eqref{eq:cont}, \eqref{eq:kirch} by following the approach proposed in \cite{Caputo2014}. By using the continuity condition \eqref{eq:cont}, we can employ for simplicity the forward finite differences written on adjacent nodes (see Figure~\ref{fig:sketch1} for the illustration). Let us denote the values of the solution at neighbouring points $\x_{\,i}\,$, $i\,=\,0,\,\ldots,\,3$ as $u_{\,i}\ \eqdef\ u\,(\x_i)\,$. Then, the discrete compatibility condition reads:
\begin{equation}\label{eq:tri}
  \frac{u_{\,1}\ -\ u_{\,0}}{\Delta x}\ +\ \frac{u_{\,2}\ -\ u_{\,0}}{\Delta x}\ +\ \frac{u_{\,3}\ -\ u_{\,0}}{\Delta x}\ =\ 0 \quad \Rightarrow \quad u_{\,0}\ =\ \third\,\bigl(u_{\,1}\ +\ u_{\,2}\ +\ u_{\,3}\bigr)\,.
\end{equation}
Obviously, as in the continuous case, the last discrete condition can be generalized to any finite number of adjacent strings.


\subsection{The numerical algorithm}
\label{sec:algo}

Taking into account all the information given above, we have all the elements to describe a practical implementation of the numerical algorithm. Each edge $e_{\,i}\ \in\ E$ is discretized with $n$ equally spaced points. We have in total $m\ =\ \abs{E}$ edges. Thus, it seems natural to keep the discrete solution using three matrices $U_{\,0}\,$, $U_{\,1}\,$, $U_{\,2} \ \in\ \Mat_{\,n\times m}\,(\R)\,$. The upper index $k\ \in\ \N_{\,0}$ indicates the time step number, \ie the initial condition is discretized into $U_{\,0}$ and $U_{\,1}\,$. Please, note that the number of initial (or boundary) conditions in the continuous and discrete formulations might not coincide (see \cite[Appendix~A]{Khakimzyanov2015b} for more details). In our problem two initial conditions are required by the continuous \acs{sg} equation and two initial conditions are needed by the fully discrete scheme \eqref{eq:dsg}. In the present study we choose $U_{\,0}\,$, $U_{\,1}$ to be a perfect coherent structure (\ie kink or breather) propagating in the required direction. The pseudo-code for the discrete time evolution $U_{\,k}\,$, $k\ \geq\ 2$ is given in Algorithm~\ref{alg:alg}. We first advance the bulk of the edges and then update the graph vertices where several edges begin or end. The update is done using condition \eqref{eq:tri}. A vertex is associated to $U_{\,k}\,(1,\,j)$ if branch $j$ is out-going from it and $U_{\,k}\,(n,\,j)$ if branch $j$ is entering it. The source code, implemented in the \textsc{Matlab} environment is freely available to consult and download at the following URL: \\

\smallskip
  \url{https://github.com/dutykh/sineGordonGraph/}\bigskip

\begin{algorithm}[H]
  \centering
  \bigskip
  \begin{algorithmic}[1]
    \Require{$d\,(n_{\,d})$ } \Comment{Degree of each node}
    \Require{$\mathrm{In}\,(n_{\,d}),\, \mathrm{Out}\,(n_{\,d})$} \Comment{Input and output branches of each node}
    \Require{$U_{\,0}\,(n,\,m)$ and $U_{\,1}\,(n,\,m)$} \Comment{Initial conditions}
    \State $t\ \gets\ 0$ \Comment{We start the simulation at $t\ =\ 0$}
    \While{$t\ <\ T_{\,f}$} \Comment{$T_{\,f}$ is the final simulation time and $t$ is the current time}
    \For{$j\ \gets\ 1,\;m$} \Comment{Loop over the edges}
      \State $U_{\,2}\,(2:n-1,\,j)\ \gets\ 2\,U_{\,1}\,(2:n-1,\,j)\ -\ U_{\,0}\,(2:n-1,\,j)\ +\ \Bigl(\,\dfrac{\Delta t}{\Delta x}\,\Bigr)^2\;\L\cdot U_{\,1}\,(2:n-1,\,j)\ -\ \Delta t^{\,2}\,\sin U_{\,1}\,(2:n-1,\,j)$ \Comment{Update the solution in bulk of edges}
    \EndFor 
       \For{$\mathrm{in}\ \gets\ 1,\;n_{\,d}$} \Comment{Loop over the vertices}
      \State  $V \gets\ (\sum(U_{\,2}\,(n-1,\,\mathrm{In}\,(\mathrm{in}))) + \sum(U_{\,2}\,(n-1,\,\mathrm{Out}\,(\mathrm{in}))))/d\,(\mathrm{in})$ \Comment{Condition \eqref{eq:tri} }
      \State $U_{\,2}\,(n,\,\mathrm{In}(\mathrm{in}))\ \gets\ V\,; \quad U_{\,2}\,(1,\,\mathrm{Out}\,(\mathrm{in}))\ \gets\ V$ \Comment{Update edges of branches}
    \EndFor
    \State $U_{\,0}\,(:\,,\,:)\ \gets\ U_{\,1}\,(:\,,\,:)$
    \State $U_{\,1}\,(:\,,\,:)\ \gets\ U_{\,2}\,(:\,,\,:)$
    \State $t\ \gets\ t\ +\ \Delta t$ \Comment{Update the time variable}
    \EndWhile \Comment{End of main loop in time}
  \end{algorithmic}
  \caption{\small\em Algorithm to simulate a \textsc{Hamiltonian} equation, like the \acf{sg} equation on a network. Here $n_{\,d}$ is the degree of each vertex in the graph $G\,$. This parameter is different from $n\,$, the number of discretization points of graph edges.}
  \bigskip
  \label{alg:alg}
\end{algorithm}

\begin{figure}
  \centering
  \includegraphics[width=0.65\textwidth]{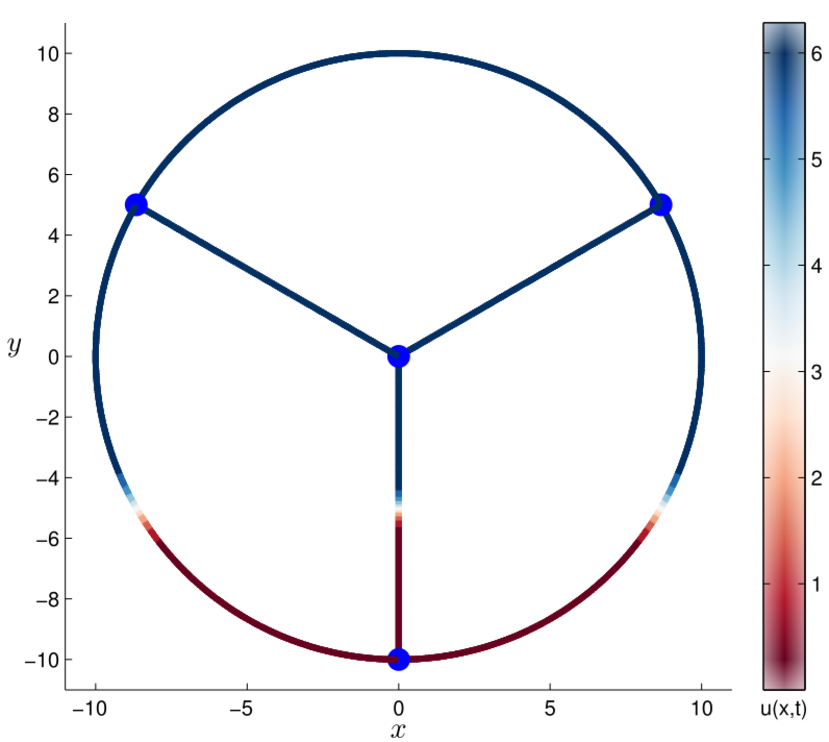}
  \caption{\small\em Embedding of the graph $G_0$ in the plane $\R^2$ along with the initial condition (represented with the color of the edges $E$) for three kinks propagating upwards in the network. Graph vertices $V$ are represented with bold blue points.}
  \label{fig:init}
\end{figure}

Note that the `bulk' advances in the edges, steps $3\,$, $4$ and $5$ can be done in parallel. Then threads need to be synchronized for the vertex update.  In order to represent graphically the solution, one has to specify also an embedding of the graph $G$ on $\R^{\,2}$ (for planar graphs and $\R^{\,3}$ in the general case), \ie a family of regular maps $g_{\,i}:\ e_{\,i}\ \in\ E\ \mapsto\ \R^{\,2}$ which satisfy the natural compatibility conditions at the vertices. For the graph $G_{\,0}$ we chose a natural embedding shown on Figure~\ref{fig:init}.


\section{Numerical results}
\label{sec:num}

Below we present several applications of the proposed numerical scheme on a particular graph. As illustrations, we present the propagation of two coherent structures (kinks and breathers) over this graph.


\subsection{Propagation of kinks}

The initial condition consists of three kinks \eqref{eq:kink0} with velocity $c\ =\ 0.95$ initially placed on the edges $e_1\,$, $e_5\,$, $e_6$ propagating vertically upwards and connecting $0$ to $2\pi\,$. Solution values on other edges are chosen in order to satisfy the continuity condition \eqref{eq:cont}. The values of all numerical parameters are given in Table~\ref{tab:params}. The total energy of this system is equal to
\begin{equation*}
  \E(0)\ =\ 3\times 8\gamma\ \approx\ 76.86151382644181\,,
\end{equation*}
where $\gamma$ is defined in \eqref{eq:kink}. For the sake of comparison, the energy at the end of the simulation was equal to $\E(T)\ \approx\ 76.97\,$, which shows good conservative properties of the scheme (the relative error is less than $1.5\%$).

\begin{table}
  \centering
  \begin{tabular}{||>{\columncolor[gray]{0.85}}c||>{\columncolor[gray]{0.85}}c||}
  \hline\hline
  \textit{Parameter} & \textit{Value} \\
  \hline\hline
    Kink speed, $c_{\,0}$ & $0.95$; $0.5$ \\
    Time step, $\Delta t$ & $0.01$ \\
    Final simulation time, $T$ & $33.0;\ 40.0$ \\
    Number of time steps, $N_{\,t}$ & $4\,000$ \\
    Number of points, $N$ & $500$ \\
    Spatial discretization step, $\Delta x$ & $0.02$ \\
  \hline\hline
  \end{tabular}
  \bigskip
  \caption{\small\em Parameters used in numerical simulations presented in this manuscript. The kink speeds are chosen in order to illustrate the phenomena of transition/reflection through/at the junction. The number of time steps is chosen to have sufficient accuracy by verifying the stability conditions.}
  \label{tab:params}
\end{table}

\begin{figure}
  \centering
  \subfigure[$t\ =\ 6.4$]{
  \includegraphics[width=0.48\textwidth]{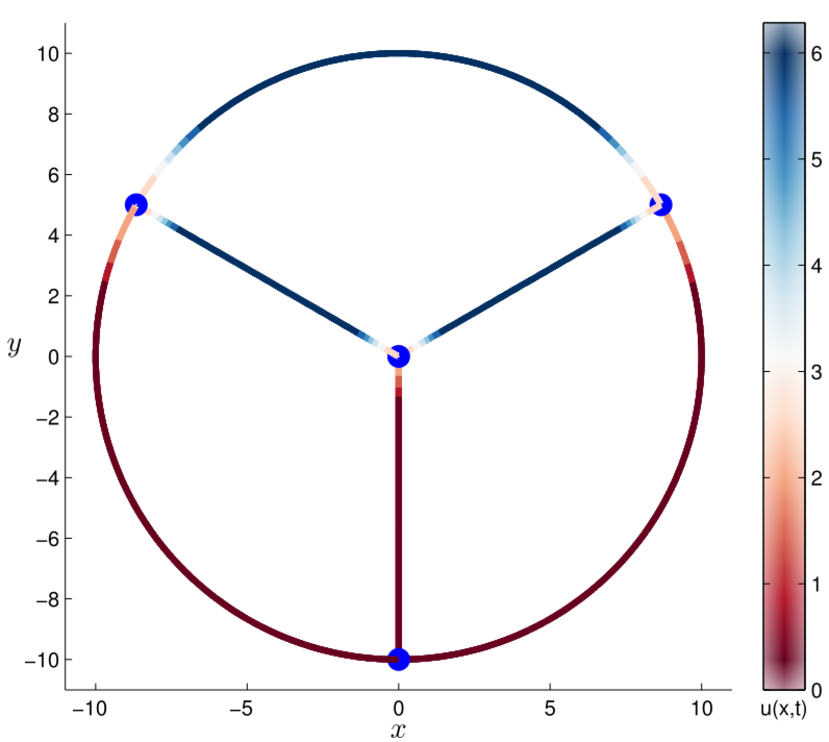}}
  \subfigure[$t\ =\ 9.45$]{
  \includegraphics[width=0.48\textwidth]{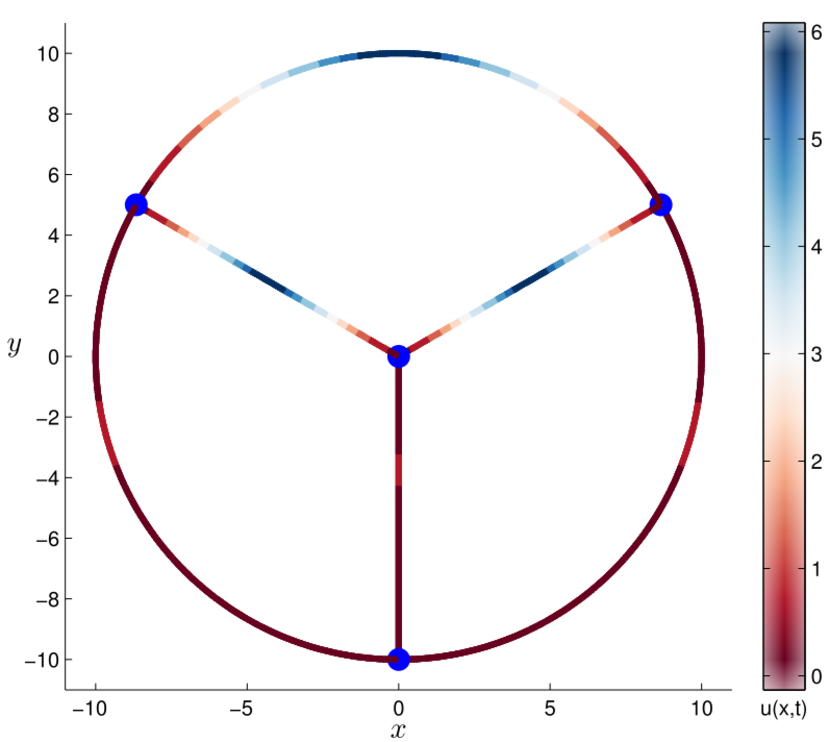}}
  \subfigure[$t\ =\ 16.3$]{
  \includegraphics[width=0.48\textwidth]{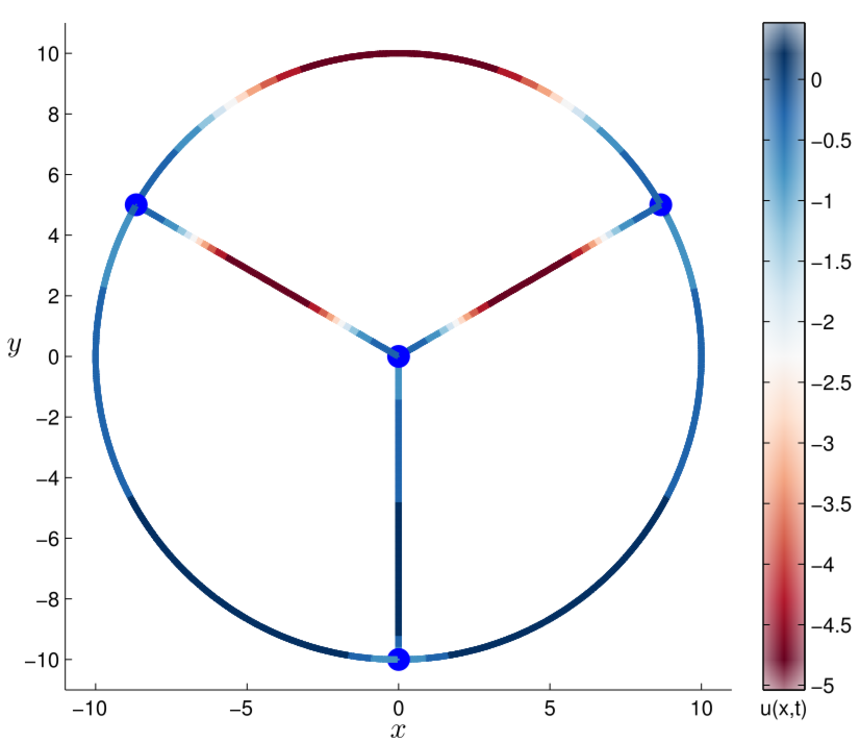}}
  \subfigure[$t\ =\ 19.2$]{
  \includegraphics[width=0.48\textwidth]{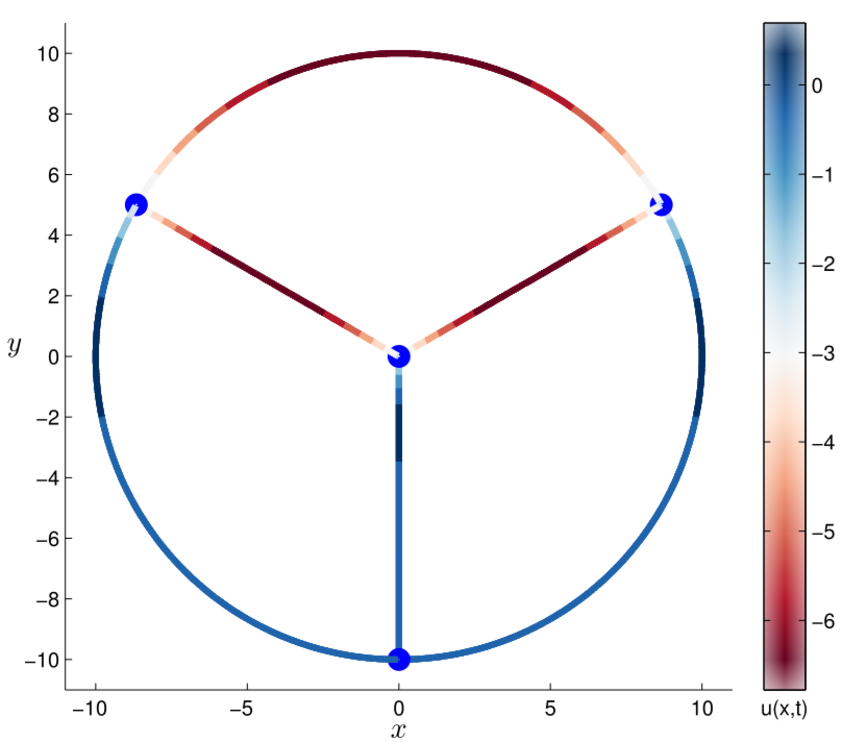}}
  \subfigure[$t\ =\ 26.1$]{
  \includegraphics[width=0.48\textwidth]{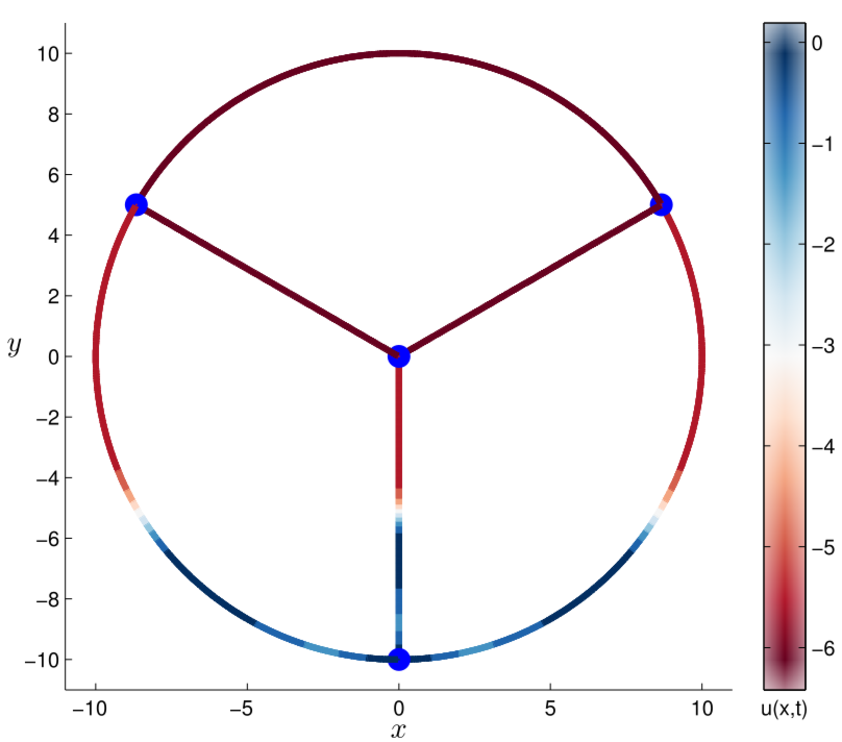}}
  \subfigure[$t\ =\ 33.0$]{
  \includegraphics[width=0.48\textwidth]{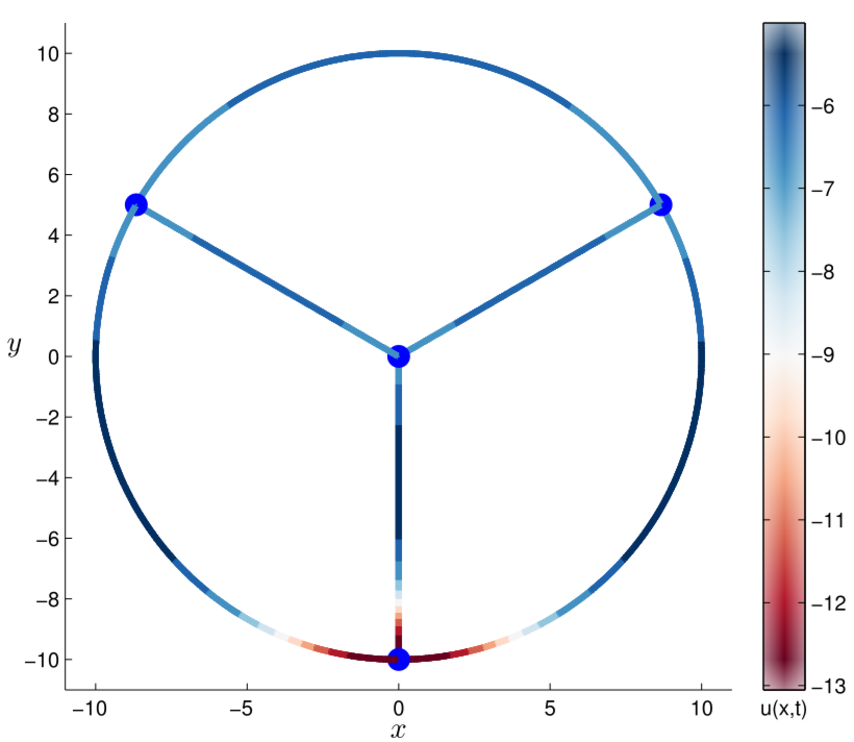}}
  \caption{\small\em Evolution of three \acs{sg} kinks (with $c\ =\ 0.95$) on the graph $G_0\,$.}
  \label{fig:dyn}
\end{figure}

The evolution of this initial condition on the interval $[\,0,\, T\,]$ under the \acs{sg} dynamics is shown on Figure~\ref{fig:dyn}. When the kinks arrive at vertices $v_{\,2}\,$, $v_{\,3}$ and $v_{\,4}$ (see Figure~\ref{fig:dyn}(\textit{a})) they split in six kinks (see Figure~\ref{fig:dyn}(\textit{b})), which collide again right in the middle of the edges $e_{\,2}\,$, $e_{\,3}$ and $e_{\,4}\,$. We observe a topological change at the moment of the collision (see Figure~\ref{fig:dyn}(\textit{c}, \textit{d})), since all the kinks switch from the ground state $0\ \leadsto\ 2\pi$ to $0\ \leadsto\ -2\pi$ as a result of the mutual reflection. Then, the newly generated kinks propagate vertically downwards along the graph edges $e_1\,$, $e_{\,5}$ and $e_{\,6}$ (see Figure~\ref{fig:dyn}(\textit{e})). Finally, at the end of the simulation the three kinks collide again in the vicinity of the vertex $v_{\,1}\,$. At the moment of collision there is another topological change from $0\ \leadsto\ -2\pi$ to $-2\pi\ \leadsto\ -4\pi$ (which are constant admissible solutions \eqref{eq:stat}). We would like to mention that we observe the generation of a small reflected wavelet into the incident branch when a kink arrives to a junction. However, we do not exclude a possibility that this wavelet might be a numerical artifact due to discretization and/or the implementation of junction conditions. The corresponding video illustration of this simulation can be watched at this URL:

\smallskip
  \url{http://youtu.be/rwZ4d_T7nTs}\bigskip

Let us perform another simulation, where we take the same set-up as described above, but the kinks are initialized to have the speed $c\ =\ 0.5\,$. The dynamics of this initial condition is shown on Figure~\ref{fig:dyn2}. In agreement with previous investigations (limited to only a single junction) \cite{Gulevich2008, Caputo2014} this kink does not possess enough energy to go through a junction. Consequently, the dynamics is confined only to the subgraph whose edges were initialized with kinks. Here again we observe a similar phenomenon to the previous case. When the kinks collide at the vertex $v_{\,1}$ (around $t\ =\ 30$), there is a topological change and all three solutions shift from $0\ \leadsto\ 2\,\pi$ to $2\,\pi\ \leadsto\ 4\,\pi\,$. The corresponding video illustration can be watched at this URL:

\smallskip
  \url{http://youtu.be/CtglMe0IcBk}\bigskip

\begin{figure}
  \centering
  \subfigure[$t\ =\ 2.0$]{
  \includegraphics[width=0.48\textwidth]{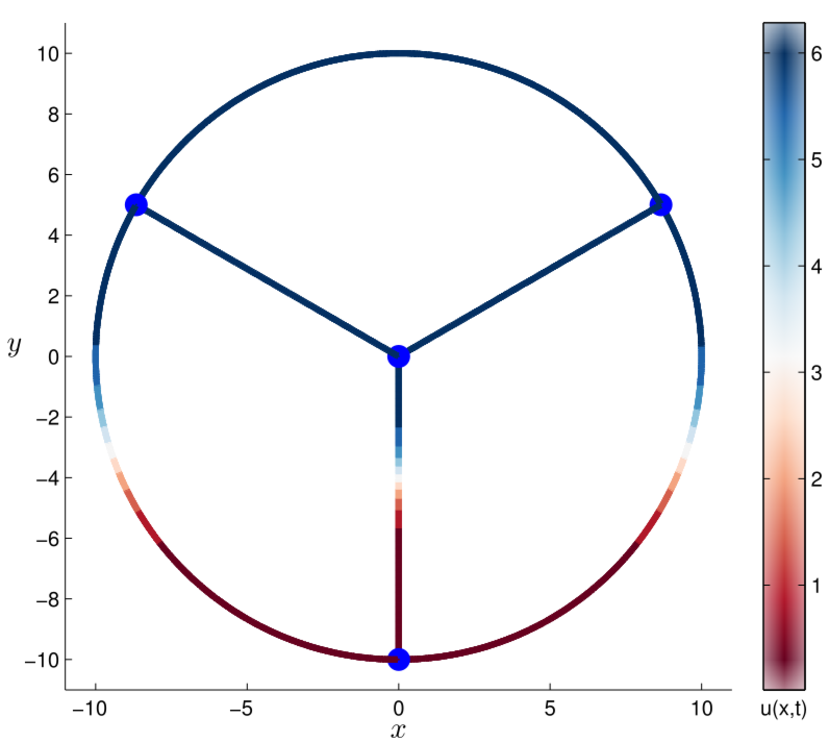}}
  \subfigure[$t\ =\ 9.0$]{
  \includegraphics[width=0.48\textwidth]{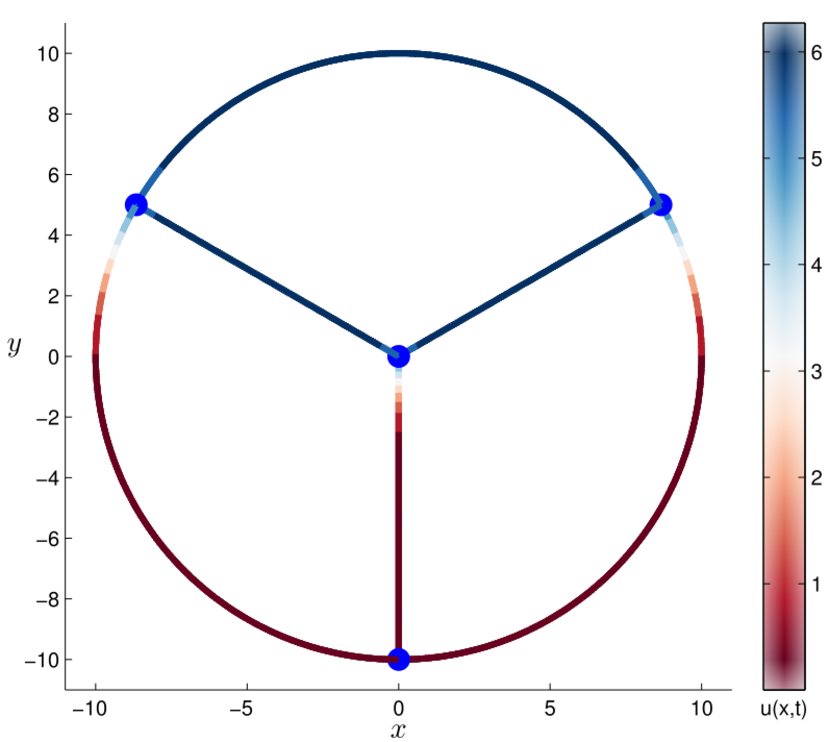}}
  \subfigure[$t\ =\ 16.0$]{
  \includegraphics[width=0.48\textwidth]{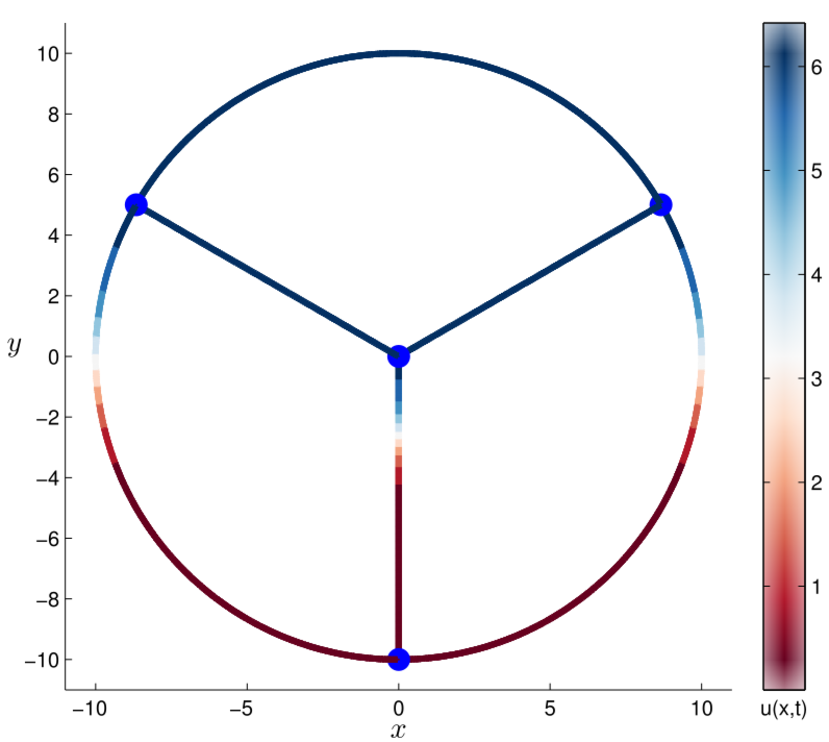}}
  \subfigure[$t\ =\ 24.0$]{
  \includegraphics[width=0.48\textwidth]{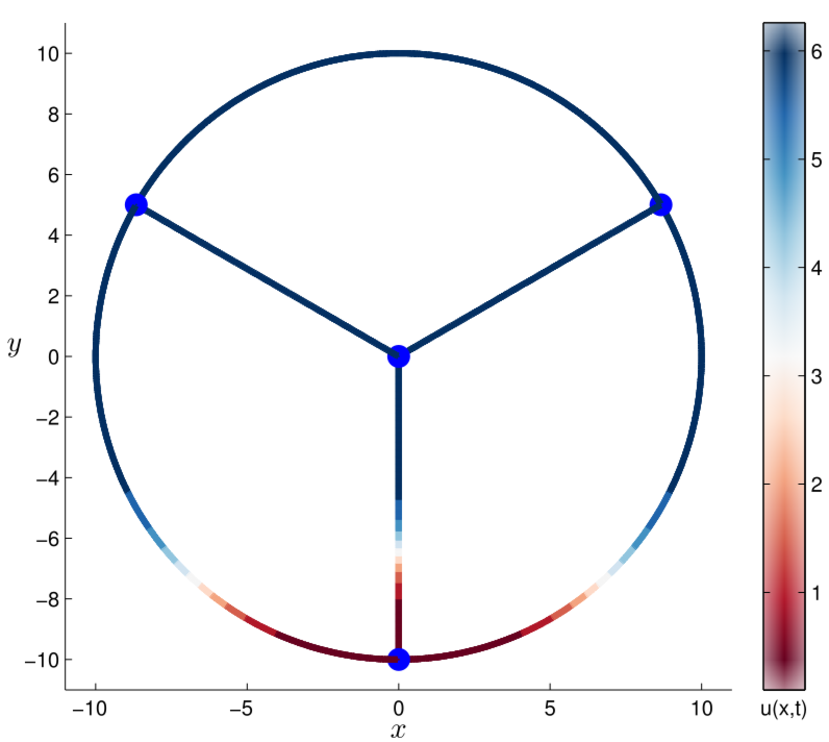}}
  \subfigure[$t\ =\ 34.0$]{
  \includegraphics[width=0.48\textwidth]{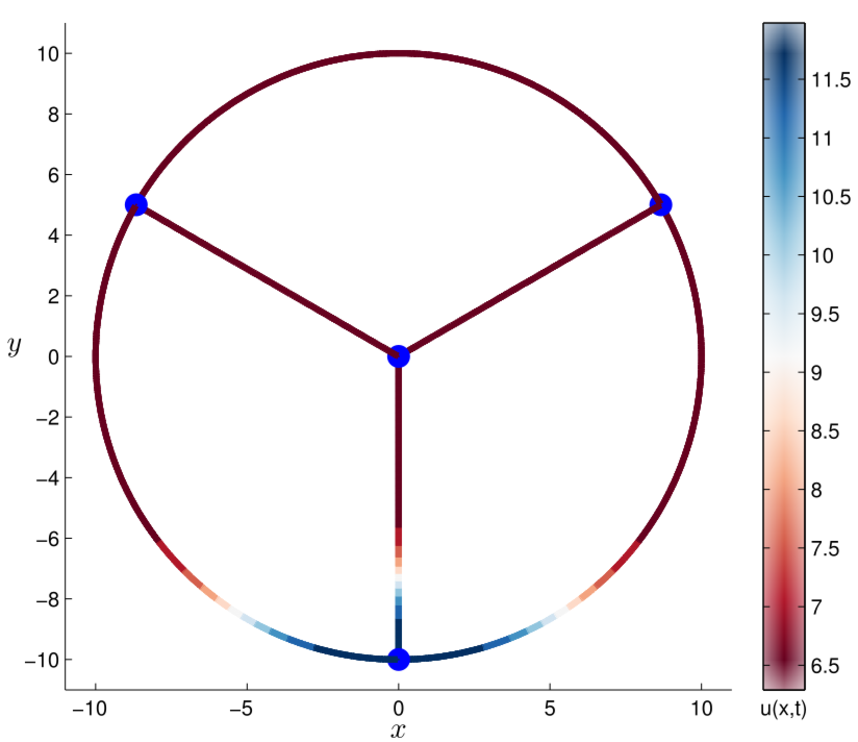}}
  \subfigure[$t\ =\ 40.0$]{
  \includegraphics[width=0.48\textwidth]{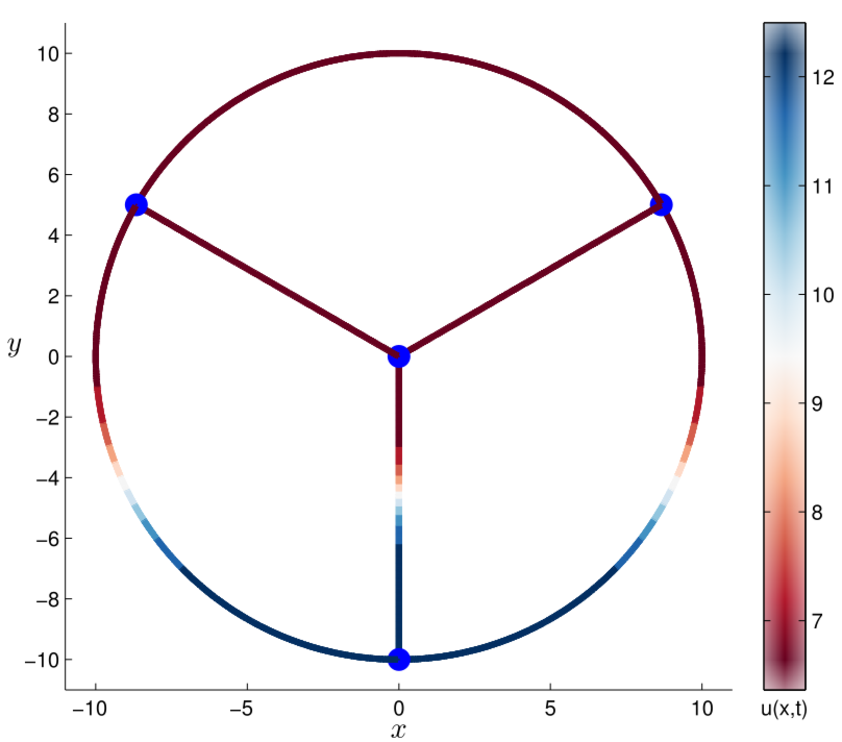}}
  \caption{\small\em Evolution of three \acs{sg} kinks (with $c_{\,0}\ =\ 0.5$) on the graph $G_{\,0}\,$.}
  \label{fig:dyn2}
\end{figure}

The presented simulations allow us to draw already some conclusions. First of all, there are two distinct situations depending on the kink energy. If this energy is sub-critical, the kinks will be confined to oscillate forever in a small part of the network where they were placed initially. The other situation with super-critical energy is much richer in terms of the generated dynamics. It seems that this dynamics will depend on the network topology. In our study we made a choice for a closed network. As a result, we observe a quasi-periodic dynamics as illustrated in Figure~\ref{fig:dyn}.


\subsection{Propagation of a breather}
\label{sec:breath}

\begin{figure}
  \centering
  \subfigure[$t\ =\ 2.0$]{
  \includegraphics[width=0.48\textwidth]{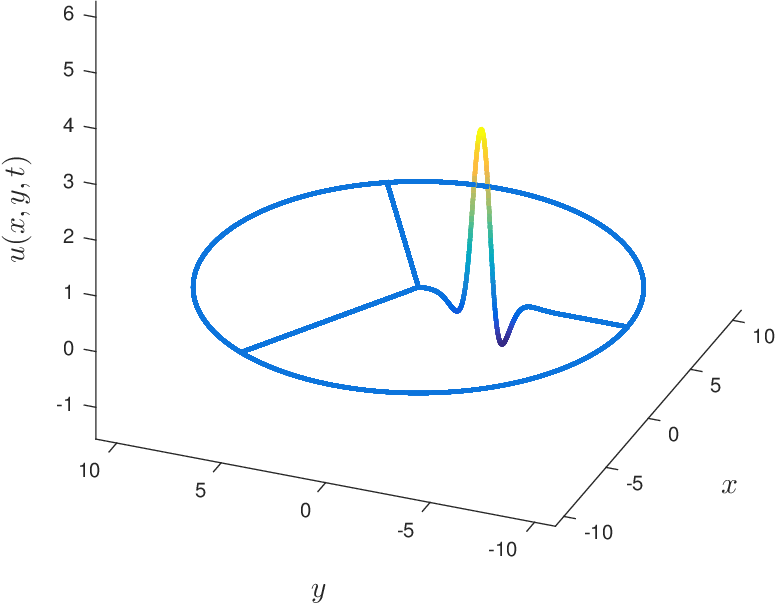}}
  \subfigure[$t\ =\ 10.0$]{
  \includegraphics[width=0.48\textwidth]{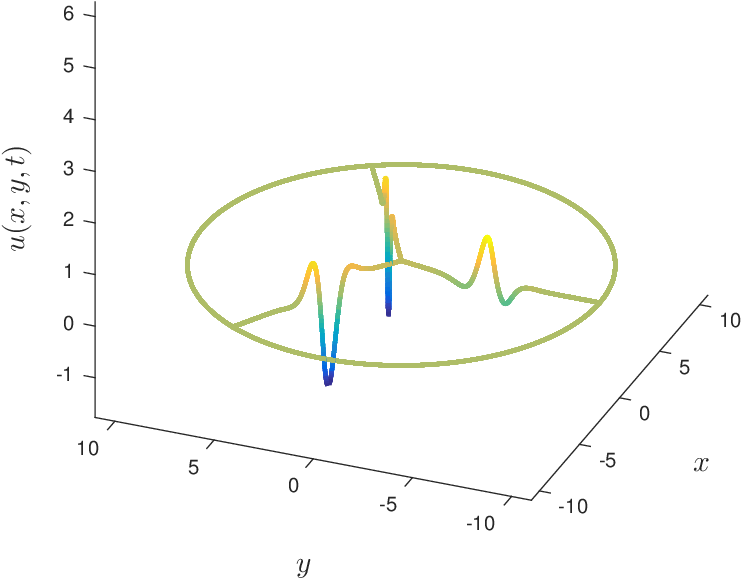}}
  \subfigure[$t\ =\ 15.0$]{
  \includegraphics[width=0.48\textwidth]{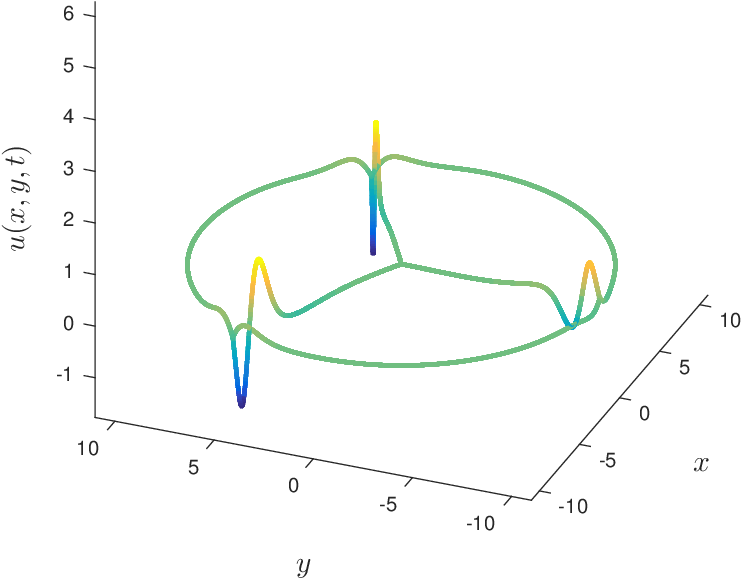}}
  \subfigure[$t\ =\ 17.0$]{
  \includegraphics[width=0.48\textwidth]{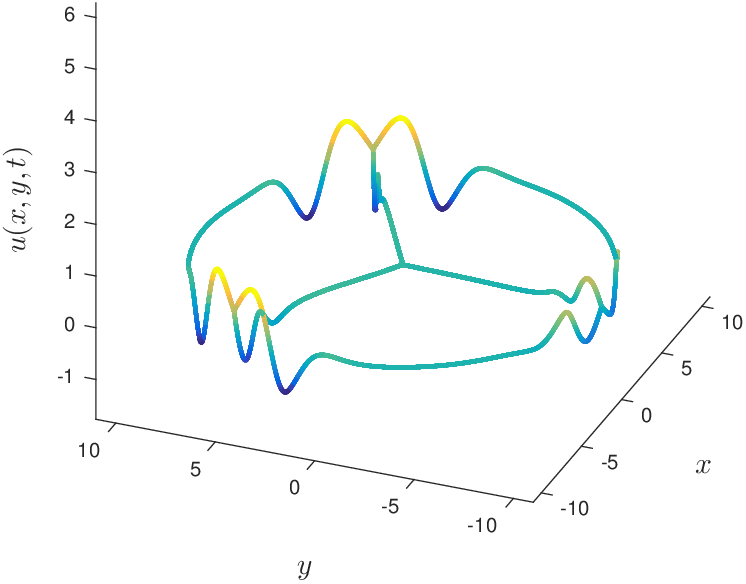}}
  \subfigure[$t\ =\ 24.0$]{
  \includegraphics[width=0.48\textwidth]{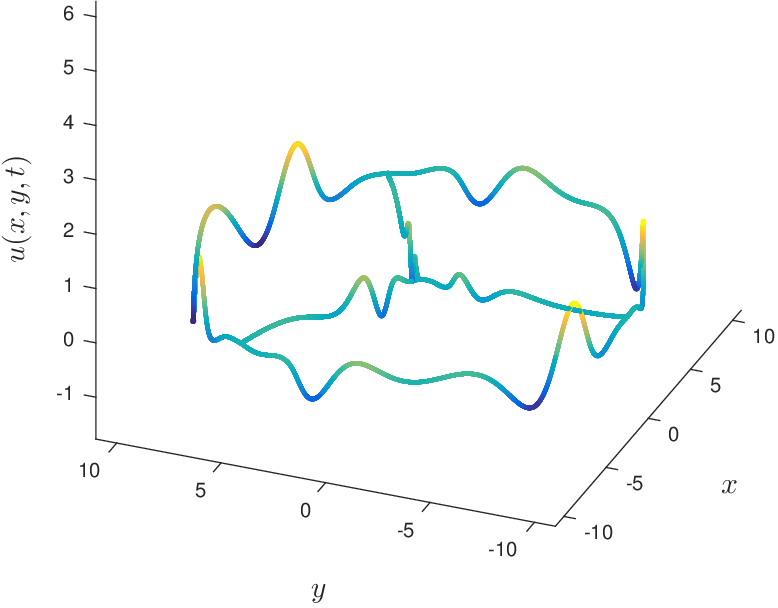}}
  \subfigure[$t\ =\ 33.0$]{
  \includegraphics[width=0.48\textwidth]{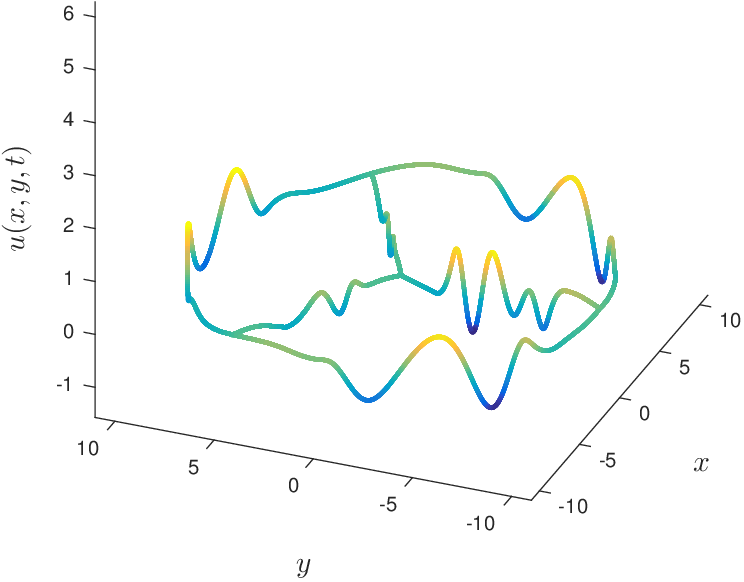}}
  \caption{\small\em Evolution of a sine--Gordon breather with $c_{\,0}\ =\ 0.95$ and $\omega\ =\ \frac{\sqrt{2}}{2}$ on the graph $G_{\,0}\,$.}
  \label{fig:dyn3}
\end{figure}

The propagation of kinks in simple junctions is becoming a relatively studied topic \cite{Hattel1996, Gulevich2006, Caputo2014}. In our study we make a further step towards numerical simulations on general networks, which contain numerous junctions. Moreover, our methodology is not limited only to kinks. To illustrate the performance of our algorithm with more general initial conditions, we complete our study with a test case of breather propagation over the same network, which is depicted in Figure~\ref{fig:merc}. The numerical (discretization) parameters and the graph embedding are taken the same as above (see Table~\ref{tab:params}). The breather has the initial speed $c_{\,0}\ =\ 0.95$ and the frequency $\omega\ =\ \cos\bigl(\frac{\pi}{4}\bigr)\ \equiv\ \dfrac{\sqrt{2}}{2}\,$. Thus, the parameter $\mu\ =\ \dfrac{\pi}{4}\,$. The energy of this breather is equal to
\begin{equation*}
  \E\,(0)\ =\ 16\,\gamma\,\sin(\mu)\ \approx\ 36.23286509262705\,.
\end{equation*}
Initially the breather is located at the edge $e_{\,1}$ and travels towards the junction point $v_{\,2}\,$ (see Figure~\ref{fig:merc}). The breather energy is conserved within $1\%$ relative accuracy along the simulation ($t\ \in\ [\,0,\,T\,]\,$, $T\ =\ 33.0$) and the total energy evolution has no trend (there are mainly oscillations around the mean level). The evolution of this breather is shown\footnote{We change the view angle in order to illustrate better the breather evolution.} in Figure~\ref{fig:dyn3}. The video of this process can be visualized also at this URL address:

\smallskip
  \url{https://youtu.be/doKNMrOkrOo}\bigskip

From this numerical simulation we can draw the following \emph{preliminary} conclusions on breather dynamics on networks:
\begin{itemize}
  \item A breather passing through a $Y-$junction point is partly reflected (see Figure~\ref{fig:dyn3}(\textit{b})). However, most of the energy is transmitted and two newly generated breathers seem to change the polarity, \ie the amplitude changes the sign. As a result, we obtain three breathers (see Figure~\ref{fig:dyn3}(\textit{c})). This property to conserve the type of the coherent structure through the junction should be reminiscent of the integrability of the \acs{sg} equation
  \item The same happens at every junction crossing event: a partial reflection\footnote{The reflected wave is a breather as well of the same polarity.} and the generation of two new breathers (see Figure~\ref{fig:dyn3}(\textit{d}))
  \item There are important differences with the propagation of kinks. Namely, the kinks pass a junction point without reflecting a breather back into the incident branch. This point is fundamental and it explains why the dynamics of kinks is easier to understand than the dynamics of a single breather in a closed network
  \item Thus, even if we start with one breather, the dynamics on a closed network becomes rapidly very complicated since the number of coherent structures might increase exponentially (\ie each crossing generates two new breathers, see Figures~\ref{fig:dyn3}(\textit{e,\,f}))
  \item While propagating in branches, the breathers interact with each other
\emph{elastically}\footnote{The term `elastic' means that coherent structures recover their initial shape after the interaction in contrast to `inelastic' collisions. The property of elasticity in interactions remains rather exceptional since integrable models are exceptional in the world of PDEs.}, since the \acs{sg} equation is integrable \cite{Takhtadzhyan1974, Dashen1974}
  \item If we had an infinite resolution, we would probably observe something similar to solitonic turbulence in \textsc{Korteweg}--\textsc{de Vries}-like models \cite{Zakharov1988, Dutykh2014d}
  \item However, the system being conservative, the total energy is constant. Thus, the amplitude of breathers can only decrease taking into account the exponential growth of their number. As a result, we deal with decreasingly smaller objects.
\end{itemize}
To conclude this Section, it would be extremely interesting to study the long time dynamics of such systems, which would require infinite numerical resolutions to capture smaller and smaller coherent structures. Thus, it has to be done theoretically and analytically in future investigations.

\subsubsection{Weak energy breather evolution}

\begin{figure}
  \centering
  \subfigure[$t\ =\ 2.0$]{
  \includegraphics[width=0.48\textwidth]{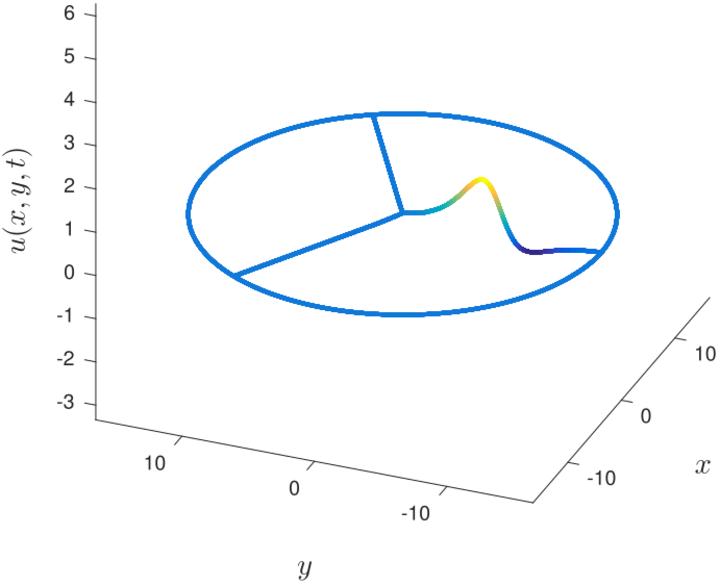}}
  \subfigure[$t\ =\ 10.0$]{
  \includegraphics[width=0.48\textwidth]{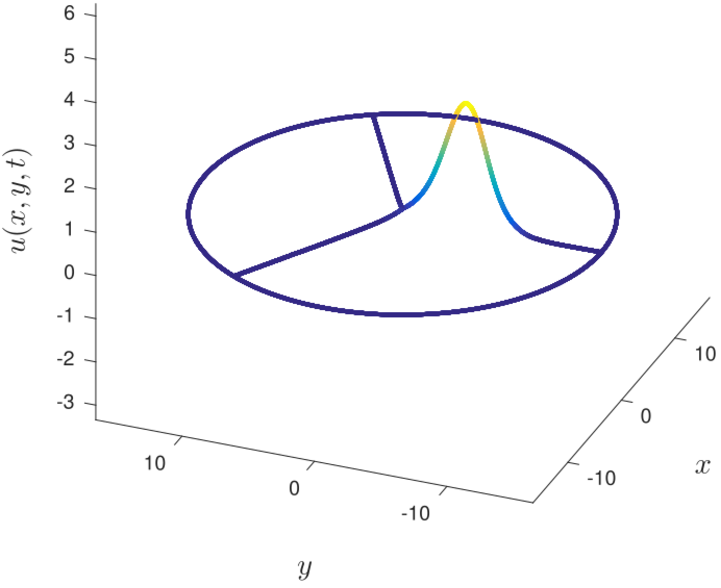}}
  \subfigure[$t\ =\ 15.0$]{
  \includegraphics[width=0.48\textwidth]{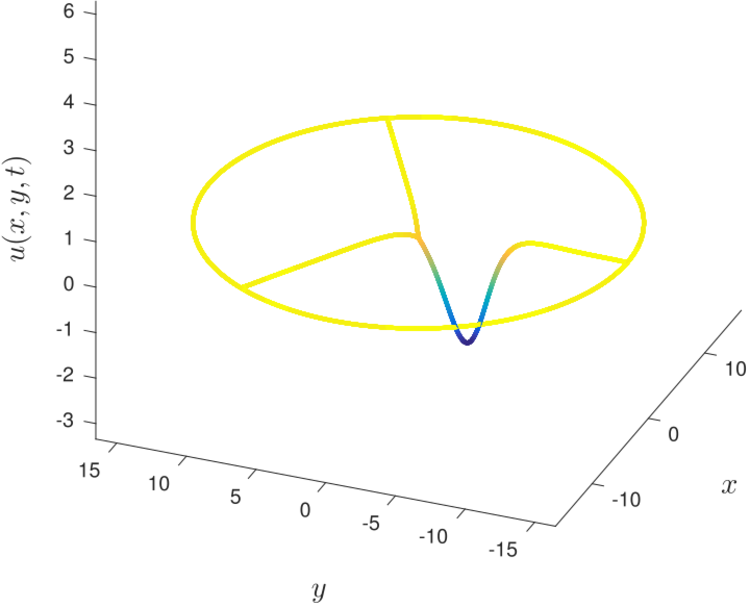}}
  \subfigure[$t\ =\ 21.0$]{
  \includegraphics[width=0.48\textwidth]{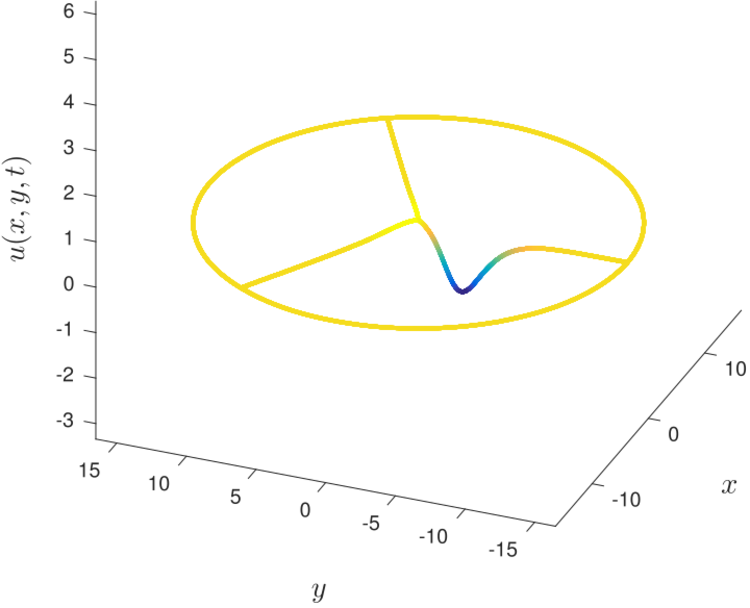}}
  \subfigure[$t\ =\ 30.0$]{
  \includegraphics[width=0.48\textwidth]{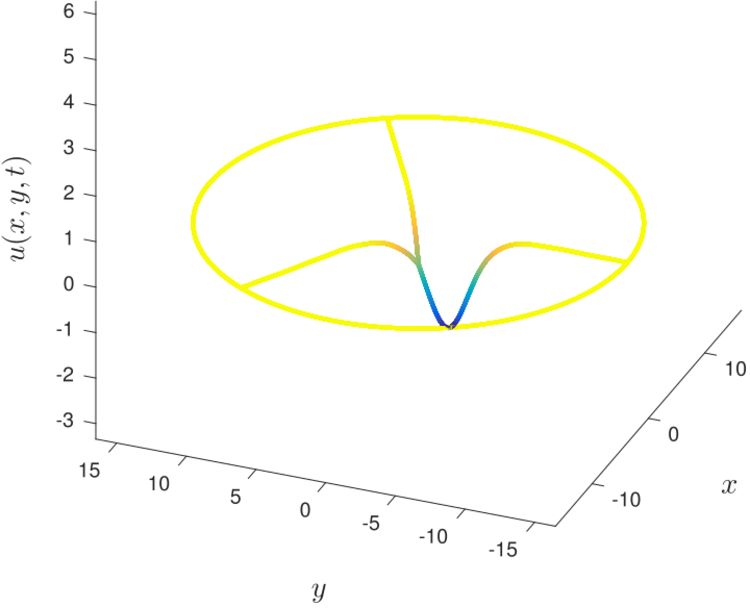}}
  \subfigure[$t\ =\ 36.0$]{
  \includegraphics[width=0.48\textwidth]{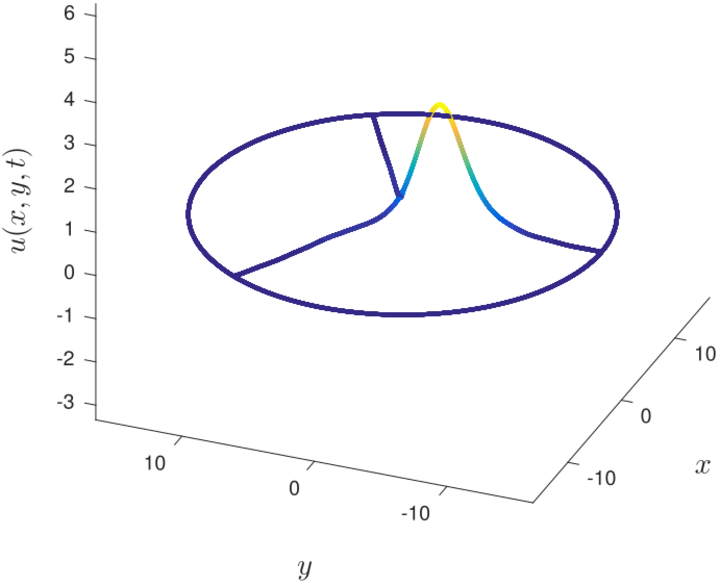}}
  \caption{\small\em Evolution of a sine--Gordon `weak' energy breather on the graph $G_{\,0}$ with parameters given in Table~\ref{tab:params2}.}
  \label{fig:dyn4}
\end{figure}

As a final test case, we consider the evolution of a breather (on the same network) with the energy below the passage barrier. The initial condition is conceptually the same as in the previous Section~\ref{sec:breath}, with different numerical parameters, which are given in Table~\ref{tab:params2}. Notice, that we had to take slightly longer branches to put this breather entirely into the edge $e_{\,1}\,$. The mesh was refined accordingly to keep approximatively the same level of accuracy. The energy of the `weak' breather is equal to
\begin{equation*}
  \E\,(0)\ =\ 16\,\gamma\,\sin(\mu)\ \approx\ 11.68474789344354\,.
\end{equation*}
This energy was conserved with the relative accuracy $<\ 1\%$ in our simulation. The evolution of this initial `weak' energy breather is shown in Figure~\ref{fig:dyn4}. One can see that the breather remains confined forever to the edge $e_{\,1}$ as expected. It is interesting to note that the breather `sticks' somehow to the junction point $v_{\,2}$ and oscillates with it. The breather is apparently never reflected by the junction. All other nodes remain unaffected. The complete video of this process can be visualized also at this URL address:

\smallskip
  \url{https://youtu.be/sncYZU-cbkY/}\bigskip

\begin{table}
  \centering
  \begin{tabular}{||>{\columncolor[gray]{0.85}}c||>{\columncolor[gray]{0.85}}c||}
  \hline\hline
  \textit{Parameter} & \textit{Value} \\
  \hline\hline
    Breather speed, $c_{\,0}$ & $0.95$; $0.25$ \\
    Breather frequency, $\cos(\mu)$ & $\frac{\sqrt{2}}{2}$, \ie $\mu\ =\ \frac{\pi}{4}$ \\
    Breather energy, $\E$ & $11.68$ \\
    Time step, $\Delta t$ & $0.0075$ \\
    Final simulation time, $T$ & $36.0$ \\
    Number of time steps, $N_{\,t}$ & $4\,800$ \\
    Number of points, $N$ & $1\,000$ \\
    Spatial discretization step, $\Delta x$ & $0.015$ \\
  \hline\hline
  \end{tabular}
  \bigskip
  \caption{\small\em Parameters used in the numerical simulation of the breather propagation on a network. The breather speeds are chosen in order to illustrate the phenomena of transition/reflection through/at the junction. The number of time steps is chosen to have sufficient accuracy by verifying the stability conditions.}
  \label{tab:params2}
\end{table}

\section{Conclusions and perspectives}
\label{sec:concl}

We considered a discrete formulation of a scalar \textsc{Hamiltonian} equation on domains which are not manifolds and applied it to the \acs{sg} equation. More precisely, the 1D lattices are assembled into arbitrary graphs (networks) and the coupled dynamics can be efficiently simulated using our methodology based on a simple symplectic numerical scheme. The edges of the graph deserve a special treatment based on local conservation laws. The performance of this formulation is illustrated on a connected graph involving four cycles. Our computational methodology can easily be generalized to the case where edges or vertices are active as for many engineering applications like fluid or traffic networks. The \textsc{Matlab} source code, which implements the algorithms described in this study can be freely accessed at the following URL address: \\

\smallskip
  \url{https://github.com/dutykh/sineGordonGraph/}\bigskip

The numerical results revealed a sequence of topological changes during the collisions of elementary \acs{sg} kinks at the graph vertices. We considered two different situations when the kinks initially were super- and sub-critical. In the latter case the system dynamics is restricted only to a sub-graph since the kinks do not possess enough energy to go through the junctions. We also studied the dynamics of a breather in such a network. To our knowledge, these results are presented for the first time and shed light on a type of soliton turbulence.


\subsection*{Acknowledgments}
\addcontentsline{toc}{subsection}{Acknowledgments}

The authors acknowledge the ``Centre de Ressources Informatiques de \textsc{Haute Normandie}'' where most of the calculations were done. Moreover, we thank Professor Michel \textsc{Raibaut} (LAMA UMR \#5127, Universit\'e \textsc{Savoie Mont Blanc}) for stimulating discussions on various geometrical matters.


\appendix
\section{Conservation laws approach}
\label{app:cons}

A weaker version of condition \eqref{eq:kirch} can be derived from variational considerations by following \cite{Bibikov2009}. Let us consider again the \textsc{Lagrangian} density \eqref{eq:ld}. By using the \textsc{Taylor} expansion for the potential energy term, the \textsc{Lagrangian} density $\L_{sg}$ can be rewritten as
\begin{equation*}
  \L_{\,\mathrm{sG}}\ =\ \underbrace{\half\,\bigl(u_t^2\ -\ u_x^2\bigr)\ -\ \half\, u^2}_{\L_{\mathrm{KG}}}\ +\ \sum_{k\,=\,2}^{\infty}(-1)^{k}\frac{u^{\,2k}}{(2k)!}\ =\ \L_{\mathrm{KG}} \ +\ \sum_{k\,=\,2}^{\infty}(-1)^{k}\frac{u^{\,2k}}{(2k)!}\,.
\end{equation*}
The last form of the \textsc{Lagrangian} density is particularly suitable for the complexification of the \acs{sg} equation which derives from the following \textsc{Lagrangian}
\begin{equation}\label{eq:Lc}
  \L^{\,c}_{\,\mathrm{sG}}\ =\ \half\,\bigl(u_t\, u^*_t\ -\ u_x\, u_x^*\bigr)\ -\ \half\, u\, u^*\ +\ \sum_{k\,=\,2}^{\infty}(-1)^{k}\frac{(u\, u^*)^k}{(2k)!}\,,
\end{equation}
where $u^*(x,\,t)$ is the complex conjugate of $u(x,t)$ (in the complex-valued version of the \acs{sg} equation).

Consider now a complex-valued field $u\,(x,\,t):\ \R\times\R^{\,+}\ \mapsto\ \C$ whose behaviour is described by \textsc{Lagrangian} density \eqref{eq:Lc}. The energy-momentum tensor $\Tt\ =\ (\T^{\,\alpha\,\beta})$ of the field $u\,(x,\,t)$ satisfies the following conservation laws
\begin{equation}\label{eq:cons}
  \partial_{\,\alpha}\T^{\,\alpha\,\beta}\ =\ 0\,,
\end{equation}
where $\partial_{\,0}\ \eqdef\ \partial_{\,t}\,$, $\partial_{\,1}\ \eqdef\ \partial_{\,x}\,$. The components of tensor $\Tt$ are given in \cite{Ryder1996}
\begin{equation}\label{eq:tensor}
  \T^{\,\alpha\,\beta}\ =\ \frac{\delta \L^{c}_{\mathrm{sG}}}{\delta (\partial_\alpha u)}\;\partial^{\,\beta} u\ +\ \frac{\delta \L^{c}_{\mathrm{sG}}}{\delta (\partial_\alpha u^*)}\;\partial^{\,\beta} u^*\ -\ g^{\alpha\beta}\L^{c}_{\mathrm{sG}}, \qquad \alpha,\; \beta\ =\ 0,\,1\,,
\end{equation}
where $g^{\,\alpha\,\beta}\ \eqdef\ \diag\{1,\, -1\}$ is the \textsc{Minkowski} tensor. The contravariant derivative operator $\partial^{\,\alpha}$ is related to $\partial_{\,\beta}$ by $\partial^{\,\alpha}\ \eqdef\ g^{\,\alpha\,\beta}\partial_{\,\beta}\,$. It is straightforward to compute the components of $\Tt$ from \eqref{eq:tensor}:
\begin{equation}\label{eq:comp}
  \T^{\,0\,0}\ =\ \half\,\partial_{\,0}\, u\,\partial_{\,0}\, u^*\ +\ \half\,\partial_{\,1}\, u\,\partial_{\,1}\, u^*\ +\ 1\ -\ \cos u, \quad \T^{\,0\,1}\ =\ -\bigl(\partial_{\,0}\,u\,\partial_{\,1}\,u^*\ +\ \partial_{\,0}\,u^*\partial_{\,1}\,u\bigr)\,.
\end{equation}
One can easily identify $\T^{\,0\,0}$ with the energy density and $\T^{\,0\,1}$ with the energy flux (see also equation \eqref{eq:cons}). Postulating the energy conservation on the $Y-$junction one obtains the following condition (see \cite{Bibikov2009} for more details):
\begin{equation*}
  \left.\T^{\,1\,0} \right\vert_{\x\ \to\ C,\; \x\ \in\ \S_{\,1}}\ +\ \left.\T^{\,1\,0} \right\vert_{\x\ \to\ C,\; \x\ \in\ \S_{\,2}}\ +\ \left.\T^{\,1\,0} \right\vert_{\x\ \to\ C,\; \x\ \in\ \S_{\,3}}\ =\ 0\,,
\end{equation*}
which can be expanded according to \eqref{eq:comp} to give finally the following analogue of the \textsc{Kirchhoff} condition:
\begin{equation}\label{eq:kirch2}
  \partial_{\,t}\,  u^*\cdot\bigl(\partial^{\,1}\,u\ +\ \partial^{\,2}\,u\ +\ \partial^{\,3}\,u\bigr)\ +\ \partial_{\,t}\, u\cdot\bigl(\partial^{\,1}\,u^*\ +\ \partial^{\,2}\,u^*\ +\ \partial^{\,3}\,u^*\bigr)\ =\ 0\ \mbox{ at }\ \x\ =\ C\,.
\end{equation}
The last condition is weaker than \eqref{eq:kirch} in the sense that all solutions to \eqref{eq:kirch} satisfy \eqref{eq:kirch2}. However, the inverse generally is not true.


\bigskip
\addcontentsline{toc}{section}{References}
\bibliographystyle{abbrv}
\bibliography{biblio}

\begin{thebibliography}{10}

\bibitem{Ablowitz1973}
M.~J. Ablowitz, D.~J. Kaup, A.~C. Newell, and H.~Segur.
\newblock {Method for Solving the Sine-Gordon Equation}.
\newblock {\em Phys. Rev. Lett}, 30(25):1262--1264, jun 1973.

\bibitem{Abraham1988}
R.~Abraham, J.~E. Marsden, and T.~Ratiu.
\newblock {\em {Manifolds, Tensor Analysis, and Applications}}, volume~75.
\newblock Springer-Verlag, New York, 1988.

\bibitem{Argyris1991}
J.~Argyris, M.~Haase, and J.~C. Heinrich.
\newblock {Finite element approximation to two-dimensional sine-Gordon
  solitons}.
\newblock {\em Comput. Methods Appl. Mech. Engrg.}, 86(1):1--26, mar 1991.

\bibitem{Arnold1997}
V.~I. Arnold.
\newblock {\em {Mathematical Methods of Classical Mechanics}}.
\newblock Springer, New York, 2nd edition, 1997.

\bibitem{Bibikov2009}
P.~N. Bibikov and L.~V. Prokhorov.
\newblock {Mechanics not on a manifold}.
\newblock {\em J. Phys. A: Math. Gen}, 42(4):045302, jan 2009.

\bibitem{Bona2008}
J.~L. Bona and R.~Cascaval.
\newblock {Nonlinear dispersive waves on trees}.
\newblock {\em Canadian Applied Mathematics Quarterly}, 16(1):1--18, 2008.

\bibitem{Bressan2014}
A.~Bressan, S.~Cani{\'{c}}, M.~Garavello, M.~Herty, and B.~Piccoli.
\newblock {Flows on networks: recent results and perspectives}.
\newblock {\em EMS Surveys in Mathematical Sciences}, 1(1):47--111, 2014.

\bibitem{Caputo2014}
J.-G. Caputo and D.~Dutykh.
\newblock {Nonlinear waves in networks: model reduction for sine-Gordon}.
\newblock {\em Phys. Rev. E}, 90:022912, 2014.

\bibitem{Caputo2016}
J.-G. Caputo, D.~Dutykh, and B.~Gleyse.
\newblock {Coupling conditions for the nonlinear shallow water equations in
  forks}.
\newblock {\em Submitted}, pages 1--24, 2017.

\bibitem{Courant1928}
R.~Courant, K.~Friedrichs, and H.~Lewy.
\newblock {{\"{U}}ber die partiellen Differenzengleichungen der mathematischen
  Physik}.
\newblock {\em Mathematische Annalen}, 100(1):32--74, 1928.

\bibitem{Dashen1974}
R.~F. Dashen, B.~Hasslacher, and A.~Neveu.
\newblock {Nonperturbative methods and extended-hadron models in field theory.
  I. Semiclassical functional methods}.
\newblock {\em Phys. Rev. D}, 10(12):4114--4129, 1974.

\bibitem{Dehghan2008}
M.~Dehghan and A.~Shokri.
\newblock {A numerical method for solution of the two-dimensional sine-Gordon
  equation using the radial basis functions}.
\newblock {\em Math. Comp. Simul.}, 79(3):700--715, dec 2008.

\bibitem{Dutykh2014d}
D.~Dutykh and E.~Pelinovsky.
\newblock {Numerical simulation of a solitonic gas in KdV and KdV-BBM
  equations}.
\newblock {\em Phys. Lett. A}, 378(42):3102--3110, aug 2014.

\bibitem{Faddeev1987}
L.~D. Faddeev and L.~Takhtajan.
\newblock {\em {Hamiltonian Methods in the Theory of Solitons}}.
\newblock Springer, Berlin Heidelberg New York, 1987.

\bibitem{Furihata2001}
D.~Furihata.
\newblock {Finite-difference schemes for nonlinear wave equation that inherit
  energy conservation property}.
\newblock {\em J. Comp. Appl. Math.}, 134(1-2):37--57, sep 2001.

\bibitem{Gould2012}
R.~Gould.
\newblock {\em {Graph Theory}}.
\newblock Dover Publications Inc., dover edition, 2012.

\bibitem{Grunnet-Jepsen1993}
A.~Grunnet-Jepsen, F.~N. Fahrendorf, S.~A. Hattel, N.~Gr{\o}nbech-Jensen, and
  M.~R. Samuelsen.
\newblock {Fluxons in three long coupled Josephson junctions}.
\newblock {\em Phys. Lett. A}, 175(2):116--120, apr 1993.

\bibitem{Gulevich2006}
D.~Gulevich and F.~Kusmartsev.
\newblock {Flux Cloning in Josephson Transmission Lines}.
\newblock {\em Phys. Rev. Lett.}, 97(1):017004, jul 2006.

\bibitem{Gulevich2008}
D.~Gulevich, F.~Kusmartsev, S.~Savel'ev, V.~Yampol'skii, and F.~Nori.
\newblock {Shape Waves in 2D Josephson Junctions: Exact Solutions and Time
  Dilation}.
\newblock {\em Phys. Rev. Lett}, 101(12):127002, sep 2008.

\bibitem{Haefliger1970}
A.~Haefliger.
\newblock {Feuilletages sur les vari{\'{e}}t{\'{e}}s ouvertes}.
\newblock {\em Topology}, 9(2):183--194, may 1970.

\bibitem{Hattel1996}
S.~A. Hattel, A.~Grunnet-Jepsen, and M.~R. Samuelsen.
\newblock {Dynamics of three coupled long Josephson junctions}.
\newblock {\em Phys. Lett. A}, 221(1-2):115--123, sep 1996.

\bibitem{Ilati2015}
M.~Ilati and M.~Dehghan.
\newblock {The use of radial basis functions (RBFs) collocation and RBF-QR
  methods for solving the coupled nonlinear sine-Gordon equations}.
\newblock {\em Engineering Analysis with Boundary Elements}, 52:99--109, mar
  2015.

\bibitem{Islas2003}
A.~L. Islas and C.~M. Schober.
\newblock {Multi-symplectic Spectral Methods for the Sine-Gordon Equation}.
\newblock In P.~M.~A. Sloot, D.~Abramson, A.~V. Bogdanov, Y.~E. Gorbachev,
  J.~J. Dongarra, and A.~Y. Zomaya, editors, {\em Computational Science - ICCS
  2003}, pages 101--110. Springer, Berlin, Heidelberg, 2003.

\bibitem{Khakimzyanov2015b}
G.~Khakimzyanov and D.~Dutykh.
\newblock {On supraconvergence phenomenon for second order centered finite
  differences on non-uniform grids}.
\newblock {\em J. Comp. Appl. Math.}, 326:1--14, dec 2017.

\bibitem{Leimkuhler2004}
B.~Leimkuhler and S.~Reich.
\newblock {\em {Simulating Hamiltonian Dynamics}}, volume~14 of {\em Cambridge
  Monographs on Applied and Computational Mathematics}.
\newblock Cambridge University Press, Cambridge, 2005.

\bibitem{Lew2003}
A.~Lew, J.~Marsden, M.~Ortiz, and M.~West.
\newblock {An overview of variational integrators}.
\newblock In {\em Finite Element Methods: 1970s and beyond (CIMNE, 2003)},
  page~18, Barcelona, Spain, 2004.

\bibitem{Marsden1998}
J.~E. Marsden, G.~W. Patrick, and S.~Shkoller.
\newblock {Multisymplectic geometry, variational integrators, and nonlinear
  PDEs}.
\newblock {\em Comm. Math. Phys.}, 199(2):351--395, 1998.

\bibitem{Mehmeti2003}
F.~A. Mehmeti and V.~R{\'{e}}gnier.
\newblock {Splitting of energy of dispersive waves in a star-shaped network}.
\newblock {\em ZAMM}, 83(2):105--118, feb 2003.

\bibitem{Moore2003}
B.~Moore and S.~Reich.
\newblock {Backward error analysis for multi-symplectic integration methods}.
\newblock {\em Numerische Mathematik}, 95(4):625--652, 2003.

\bibitem{Nakajima1978}
K.~Nakajima and Y.~Onodera.
\newblock {Logic design of Josephson network. II}.
\newblock {\em J. Appl. Phys.}, 49(5):2958, 1978.

\bibitem{Nakajima1976}
K.~Nakajima, Y.~Onodera, and Y.~Ogawa.
\newblock {Logic design of Josephson network}.
\newblock {\em J. Appl. Phys.}, 47(4):1620--1627, apr 1976.

\bibitem{Rashidinia2011}
J.~Rashidinia and R.~Mohammadi.
\newblock {Tension spline solution of nonlinear sine-Gordon equation}.
\newblock {\em Numerical Algorithms}, 56(1):129--142, jan 2011.

\bibitem{Ryder1996}
L.~H. Ryder.
\newblock {\em {Quantum Field Theory}}.
\newblock Cambridge University Press, Cambridge, 2nd edition, 1996.

\bibitem{Scott2003}
A.~Scott.
\newblock {\em {Nonlinear Science: Emergence and Dynamics of Coherent
  Structures}}.
\newblock Oxford University Press, 2nd edition, 2003.

\bibitem{Scott2004}
A.~Scott.
\newblock {\em {Encyclopedia of Nonlinear Science}}.
\newblock Routledge, New-York, 2004.

\bibitem{Susanto2005}
H.~Susanto and S.~A. van Gils.
\newblock {Existence and stability analysis of solitary waves in a tricrystal
  junction}.
\newblock {\em Phys. Lett. A}, 338(3-5):239--246, may 2005.

\bibitem{Takhtadzhyan1974}
L.~A. Takhtadzhyan and L.~D. Faddeev.
\newblock {Essentially nonlinear one-dimensional model of classical field
  theory}.
\newblock {\em Theor. Math. Phys.}, 21(2):1046--1057, 1974.

\bibitem{Vassilevskii2011}
Y.~Vassilevskii, S.~Simakov, V.~Salamatova, Y.~Ivanov, and T.~Dobroserdova.
\newblock {Numerical issues of modelling blood flow in networks of vessels with
  pathologies}.
\newblock {\em Russ. J. Numer. Anal. Math. Modelling}, 26(6):605--622, 2011.

\bibitem{Wattis1996}
J.~A.~D. Wattis.
\newblock {Variational approximations to breathers in the discrete sine-Gordon
  equation II: moving breathers and Peierls-Nabarro energies}.
\newblock {\em Nonlinearity}, 9(6):1583--1598, nov 1996.

\bibitem{Zakharov1988}
V.~E. Zakharov, A.~N. Pushkarev, V.~F. Shvets, and V.~V. Yankov.
\newblock {Soliton turbulence}.
\newblock {\em JETP Lett.}, 48(2):79--82, 1988.

\bibitem{Zuazua2013}
E.~Zuazua.
\newblock {Control and stabilization of waves on 1-d networks}.
\newblock {\em Lecture Notes in Mathematics}, 2062:463--493, 2013.

\end{thebibliography}
\bigskip

\end{document}